\documentstyle[12pt]{article}
\makeatletter

\def\compoundrel#1\over#2{\mathpalette\compoundreL{{#1}\over{#2}}}
\def\compoundreL#1#2{\compoundREL#1#2}
\def\compoundREL#1#2\over#3{\mathrel
         {\vcenter{\hbox{$\m@th\buildrel{#1#2}\over{#1#3}$}}}}
\makeatother

\makeatletter
\begin{document}

\begin{flushright}
{ GUTPA/01/10/03 }
\end{flushright}
%\vskip .01in

\begin{center}
{\Large {\bf Minimal Mixing of Quarks and Leptons\\[0pt]
in the SU(3) Theory of Flavour}}

%\vspace{36pt}
\vspace{16pt}

{\bf J.L. Chkareuli} \vspace{3pt}

{\em Institute of Physics, Georgian Academy of Sciences, 380077 Tbilisi,
Georgia}

%\vspace{20pt}
\vspace{10pt}

{\bf C.D. Froggatt}

\vspace{3pt}

{\em Department of Physics and Astronomy\\[0pt]
Glasgow University, Glasgow G12 8QQ, Scotland}

%\vspace{20pt}
\vspace{10pt}

{\bf H.B. Nielsen} %\thanksref{HBN}}
\vspace{3pt}

{\em Niels Bohr Institute, Blegdamsvej 17-21, DK 2100 Copenhagen, Denmark}

\bigskip

\bigskip

{\large {\bf Abstract}}
\end{center}

We argue that flavour mixing, both in the quark and lepton sector, follows
the minimal mixing pattern, according to which the whole of this mixing is
basically determined by the physical mass generation for the first family of
fermions. So, in the chiral symmetry limit when the masses of the lightest ($%
u$ and $d$) quarks vanish, all the quark mixing angles vanish. This minimal
pattern is shown to fit extremely well the already established CKM matrix
elements and to give fairly distinctive predictions for the as yet poorly
known ones. Remarkably, together with generically small quark mixing, it
also leads to large neutrino mixing, provided that neutrino masses appear
through the ordinary ``see-saw'' mechanism. It is natural to think that this
minimal flavour mixing pattern presupposes some underlying family symmetry,
treating families of quarks and leptons in a special way. Indeed, we have
found a local chiral $SU(3)_{F}$ family symmetry model which leads, through
its dominant symmetry breaking vacuum configuration, to a natural
realization of the proposed minimal mechanism. It can also naturally
generate the quark and lepton mass hierarchies. Furthermore spontaneous CP
violation is possible, leading to a maximal CP violating phase $\delta =%
\frac{\pi }{2}$, in the framework of the MSSM extended by a high-scale $%
SU(3)_{F}$ chiral family symmetry.

\thispagestyle{empty} \newpage

\section{Introduction}

The flavour mixing of quarks and leptons is certainly one of the major
problems that presently confront particle physics (for a recent review see 
\cite{review}). Many attempts have been made to interpret the pattern of
this mixing in terms of various family symmetries---discrete or continuous,
global or local. Among them, the abelian $U(1)$ \cite{fn1,leurer,ibanezross}
and/or non-abelian $SU(2)$ \cite{volkas,dine,barbieri} and $SU(3)$ \cite
{jon,gerry} family symmetries seem the most promising. They provide some
guidance to the expected hierarchy between the elements of the quark-lepton
mass matrices and to the presence of texture zeros \cite{rrr} in them,
leading to relationships between the mass and mixing parameters. In the
framework of the supersymmetric Standard Model, such a family symmetry
should at the same time provide an almost uniform mass spectrum for the
superpartners, with a high degree of flavour conservation \cite{nilles},
that makes its existence even more necessary in the SUSY case.

Despite some progress in understanding the flavour mixing problem, one has
the uneasy feeling that, in many cases, the problem seems just to be
transferred from one place to another. The peculiar quark-lepton mass
hierarchy is replaced by a peculiar set of $U(1)$ flavour charges or a
peculiar hierarchy of Higgs field VEVs in the non-abelian symmetry case. As
a result there are not so many distinctive and testable generic predictions,
strictly relating the flavour mixing angles to the quark-lepton masses.

A commonly accepted framework for discussing the flavour problem is based on
the picture that, in the absence of flavour mixing, only the particles
belonging to the third generation $t$, $b$ and $\tau $ have non-zero masses.
All other (physical) masses and the mixing angles then appear as a result of
the tree-level mixings of families, related to some underlying family
symmetry breaking. They might be proportional to powers of some small
parameter, which are determined by the dimensions of the family symmetry
allowed operators that generate the effective (diagonal and off-diagonal)
Yukawa couplings for the lighter families in the framework of the (ordinary
or supersymmetric) Standard Model.

Recently, a new mechanism of flavour mixing, which we call Lightest Family
Mass Generation (LFMG), was proposed \cite{lfm} (see the papers in ref.~\cite
{refs} for further discussion). According to LFMG the whole of flavour
mixing for quarks is basically determined by the mechanism responsible for
generating the physical masses of the {\it up} and {\it down} quarks, $m_{u}$
and $m_{d}$ respectively. So, in the chiral symmetry limit when $m_{u}$ and $%
m_{d}$ vanish, all the quark mixing angles vanish. This, in fact minimal,
flavour mixing model was found to fit extremely well the already established
CKM matrix elements and give fairly distinctive predictions for the yet
poorly known ones.

By its nature, the LFMG mechanism is not dependent on the number of
quark-lepton families nor on any ``vertical'' symmetry structure, unifying
quarks and leptons inside a family as in Grand Unified Theories (GUTs). 
%Being completely determined by the minimality condition 
The basic LFMG leads to mass-matrices whose non-zero diagonal and
off-diagonal elements for the $N$ family case are related as follows: 
\begin{equation}
M_{22}:M_{33}:...:M_{NN}=\left| M_{12}\right| ^{2}:\left| M_{23}\right|
^{2}:...:\left| M_{N-1\ N}\right| ^{2}  \label{prop}
\end{equation}
This shows clearly that the heavier families (4th, 5th, ...), had they
existed, would be more and more decoupled from the lighter ones and from
each other. Indeed, this behaviour can to some extent actually be seen in
the presently observed CKM matrix elements involving the third family quarks 
$t$ and $b$.

The above condition (\ref{prop}) is suggestive of some underlying flavour
symmetry, probably non-abelian rather than abelian, treating families in a
special way. Indeed, for the observed three-family case, we show that the
local family $SU(3)_{F}$ symmetry \cite{jon} treating the quark and lepton
families as fundamental chiral triplets, with an appropriate dominant
symmetry breaking vacuum configuration, could lead to a natural realization
of the LFMG mechanism. Other outstanding problems in flavour physics are
also addressed within the framework of the Minimal Supersymmetric Standard
Model (MSSM) appropriately extended by a high-scale $SU(3)_{F}$ chiral
family symmetry: the quark and lepton mass hierarchies, neutrino masses and
oscillations and the possibility of spontaneous CP violation.

The paper is organized in the following way. In section 2 we give a general
exposition of the LFMG mechanism involving two alternative scenarios. In
section 3 we present the $SU(3)_{F}$ theory of flavour, treating the quark
and lepton families as fundamental chiral triplets. We show that there is a
dominant symmetry breaking vacuum configuration related with the basic
horizontal Higgs supermultiplets, triplets and sextets of $SU(3)_{F}$, which
leads to a natural realization of the LFMG mechanism. While we mainly
consider the supersymmetric case---just the MSSM---most of our argumentation
remains valid for the Standard Model (and ordinary GUTs) as well. Finally,
in section 4, we give our summary.

\section{Minimal Flavour Mixing: the Model}

\subsection{Quark Mixing}

The proposed flavour mixing mechanism, driven solely by the generation of
the lightest family mass, could actually be realized in two generic ways.

The first basic alternative (I) is when the lightest family mass ($m_{u}$ or 
$m_{d} $) appears as a result of the complex flavour mixing of all three
families. It ``runs along the main diagonal'' of the corresponding $3\times
3 $ mass matrix $M$, from the basic dominant element $M_{33}$ to the element 
$M_{22}$ (via a rotation in the 2-3 sub-block of $M$) and then to the
primordially texture zero element $M_{11}$ (via a rotation in the 1-2
sub-block). The direct flavour mixing of the first and third families of
quarks and leptons is supposed to be absent or negligibly small in $M$.

The second alternative (II), on the contrary, presupposes direct flavour
mixing of just the first and third families. There is no involvement of the
second family in the mixing. In this case, the lightest mass appears in the
primordially texture zero $M_{11}$ element ``walking round the corner'' (via
a rotation in the 1-3 sub-block of the mass matrix $M$). Certainly, this
second version of the LFMG mechanism cannot be used for both the up and the
down quark families simultaneously, since mixing with the second family
members is a basic part of the CKM phenomenology (Cabibbo mixing, non-zero $%
V_{cb}$ element, CP violation). However, the alternative II could work for
the up quark family provided that the down quarks follow the alternative I.

So, there are two possible scenarios for the LFMG mechanism to be considered.

\subsubsection{Scenario A: ``both of the lightest quark masses $m_{u}$ and $%
m_{d}$ running along the diagonal''}

We propose that both mass matrices for the Dirac fermions --- the up quarks (%
$U$ = $u$, $c$, $t$) and the down quarks ($D$ = $d$, $s$, $b$) --- in the
Standard Model, or supersymmetric Standard Model, are Hermitian with three
texture zeros of the following form:

\begin{equation}
M_{i}=\pmatrix{ 0 & a_i & 0 \cr a_i^{\ast} & A_i & b_i \cr 0 & b_i^{\ast}&
B_i \cr}\qquad i=U,\ D  \label{LFM1}
\end{equation}
It is, of course, necessary to assume some hierarchy between the elements,
which we take to be: $B_{i}\gg A_{i}\sim \left| b_{i}\right| \gg \left|
a_{i}\right| $. We derive this hierarchical structure from our $SU(3)_F$
family symmetry model in section 3.3. The zeros in the $\left( M_{i}\right)
_{11}$ elements correspond to our, and the commonly accepted, conjecture
that the lightest family masses appear as a direct result of flavour
mixings. The zeros in $\left( M_{i}\right) _{13}$ mean that only minimal
``nearest neighbour'' interactions occur, giving a tridiagonal matrix
structure.

Now our main hypothesis, that the second and third family diagonal mass
matrix elements are practically the same in the gauge and physical
quark-lepton bases, means that : 
\begin{equation}
B_{i}=(m_{t},\ m_{b})+\delta _{i}\qquad A_{i}=(m_{c},\ m_{s})+
\delta_{i}^{\prime }  \label{BA}
\end{equation}
The components $\delta _{i}$ and $\delta _{i}^{\prime }$ are supposed to be
much less than the masses of the particles in the next lightest family,
meaning the second and first families respectively: 
\begin{equation}
|\delta _{i}|\ll (m_{c},\ m_{s}) \qquad |\delta _{i}^{\prime }|\ll (m_{u},\
m_{d})  \label{deldelp}
\end{equation}
Since the trace and determinant of the Hermitian matrix $M_{i}$ gives the
sum and product of its eigenvalues, it follows that 
\begin{equation}
\delta _{i}\simeq -(m_{u},\ m_{d})  \label{del}
\end{equation}
while the $\delta _{i}^{\prime }$ are vanishingly small and can be neglected
in further considerations.

It may easily be shown that our hypothesis and related equations (\ref{BA} - 
\ref{del}) are entirely equivalent to the condition that the diagonal
elements ($A_{i}$, $B_{i}$), of the mass matrices $M_{i}$, are proportional
to the modulus square of the off-diagonal elements ($a_{i}$, $b_{i}$): 
\begin{equation}
\frac{A_{i}}{B_{i}}=\left| \frac{a_{i}}{b_{i}}\right| ^{2}\qquad i=U,\ D
\label{ABab}
\end{equation}

For the moment we leave aside the question of deriving this proportionality
condition, Eq.~(\ref{ABab}), from some underlying theory beyond the Standard
Model (see section 3) and proceed to calculate expressions for all the
elements of the matrices $M_{i}$ and the corresponding CKM quark mixing
matrix, in terms of the physical masses.

Using the conservation of the trace, determinant and sum of principal minors
of the Hermitian matrices $M_{i}$ under unitary transformations, we are led
to a complete determination of the moduli of all their elements. The results
can be expressed to high accuracy as follows: 
\begin{equation}
A_{i}=(m_{c},\ m_{s}),\qquad B_{i}=(m_{t}-m_{u},\ m_{b}-m_{d}),
\end{equation}
\begin{equation}
\left| a_{i}\right| =(\sqrt{m_{u}m_{c}},\ \sqrt{m_{d}m_{s}})
\end{equation}
\begin{equation}
\left| b_{i}\right| =(\sqrt{m_{u}m_{t}},\ \sqrt{m_{d}m_{b}})
\end{equation}
As to the CKM matrix $V$, we must first choose a parameterisation
appropriate to our picture of flavour mixing. Among many possible ones, the
original Euler parameterisation recently advocated \cite{review} is most
convenient: 
\begin{eqnarray}
V &=&\pmatrix{ c_U & s_U & 0 \cr -s_U & c_U & 0 \cr 0 & 0 & 1 \cr}\pmatrix{
e^{-i\phi} & 0 & 0 \cr 0 & c & s \cr 0 & -s & c \cr}\pmatrix{c_D & -s_D & 0
\cr s_D & c_D & 0 \cr 0 & 0 & 1 \cr} \\
&=&\pmatrix{ s_Us_Dc + c_Uc_D e^{-i\phi} & s_Uc_Dc - c_Us_D e^{-i\phi} &
s_Us \cr c_Us_Dc - s_Uc_D e^{-i\phi} & c_Uc_Dc + s_Us_D e^{-i\phi} & c_Us
\cr -s_Ds & -c_Ds & c \cr}
\end{eqnarray}
Here $s_{U,\ D}\equiv \sin \theta _{U,\ D}$ and $c_{U,\ D}\equiv \cos \theta
_{U,\ D}$ parameterise simple rotations $R_{12}^{U,\ D}$ between the first
and second families for the up and down quarks respectively, while $s\equiv
\sin \theta $ and $c\equiv \cos \theta $ parameterise a rotation between the
second and third families. This representation of $V$ takes into account the
observed hierarchical structure of the quark masses and mixing angles. The
CP violating phase is connected directly to the first and second families
alone.

The quark mass matrices $M_U$ and $M_D$ are diagonalised by unitary
transformations which can be written in the form: 
\begin{equation}
V_U = R_{12}^U R_{23}^U \Phi^U \qquad V_D = R_{12}^D R_{23}^D \Phi^D
\label{VUVD}
\end{equation}
where $\Phi^{U,\ D}$ are phase matrices, depending on the phases of the
off-diagonal elements $a_i = \left| a_i \right| e^{i\alpha_i}$ and $b_i =
\left| b_i \right| e^{i\beta_i}$: 
\begin{equation}
\Phi^U = \pmatrix{e^{i\alpha_U} & 0 & 0 \cr 0 &1 & 0 \cr 0 & 0 &
e^{-i\beta_U} \cr} \qquad \Phi^D = \pmatrix{e^{i\alpha_D} & 0 & 0 \cr 0 &1 &
0 \cr 0 & 0 & e^{-i\beta_D} \cr}
\end{equation}
The CKM matrix is defined by 
\begin{equation}
V = V_U V_D^{\dagger} = R_{12}^U R_{23}^U \Phi^U \left( \Phi^D
\right)^{\ast} \left( R_{23}^D \right)^{-1} \left( R_{12}^D \right)^{-1}
\label{CKM1}
\end{equation}
and, after a suitable re-phasing of the quark fields, we can use the
representation 
\begin{equation}
R_{23}^U \Phi^U \left( \Phi^D \right)^{\ast} \left( R_{23}^D \right)^{-1} = %
\pmatrix{ e^{-i\phi} & 0 & 0 \cr 0 & c & s \cr 0 & -s & c \cr}
\end{equation}
The rotation matrices $R_{23}^{U,\ D}$ and $R_{12}^{U,\ D}$ for our mass
matrices are readily calculated and the CKM matrix expressed in terms of
quark mass ratios 
\begin{equation}
s_U = \sqrt{\frac{m_u}{m_c}} \qquad s_D = \sqrt{\frac{m_d}{m_s}}
\end{equation}
\begin{equation}
s = \left| \sqrt{\frac{m_d}{m_b}} - e^{i\gamma} \sqrt{\frac{m_u}{m_t}}
\right|  \label{sgam}
\end{equation}
and two phases $\phi = \alpha_D -\alpha_U$ and $\gamma = \beta_D - \beta_U$.

It follows that the Cabibbo mixing is given by the well-known Fritzsch
formula \cite{fritzsch2} 
\begin{equation}
\left| V_{us}\right| \simeq \left| s_{U}-s_{D}e^{-i\phi }\right| =\left| 
\sqrt{\frac{m_{d}}{m_{s}}}-e^{-i\phi }\sqrt{\frac{m_{u}}{m_{c}}}\right|
\end{equation}
which fits the experimental value well, provided that the CP violating phase 
$\phi $ is required to be close to $\frac{\pi }{2}$. So, in the following,
we shall assume maximal CP violation in the form $\phi =\frac{\pi }{2}$, as
is suggested by spontaneous CP violation in the framework of $SU(3)_{F}$
family symmetry (see section 3.3). The other phase $\gamma $ appearing in $%
V_{cb}$ and $V_{ub}$ 
\begin{equation}
\left| V_{cb}\right| \simeq s\qquad \left| V_{ub}\right| =s_{U}s
\end{equation}
can be rather arbitrary, since the contribution $\sqrt{\frac{m_{u}}{m_{t}}}$
to $s$ is relatively small, even compared with the uncertainties coming from
the light quark masses themselves. This leads to our most interesting
prediction (with the mass ratios calculated at the electroweak scale \cite
{koide}): 
\begin{equation}
\left| V_{cb}\right| \simeq \sqrt{\frac{m_{d}}{m_{b}}}=0.038\pm 0.007
\label{Vcb}
\end{equation}
in good agreement with the current data $\left| V_{cb}\right| =0.039\pm
0.003 $ \cite{data}. For definiteness we shall assume the phase $\gamma $ in 
$s$, see Eq.~(\ref{sgam}), to be aligned with the CP violating phase $\phi $%
, again as suggested by $SU(3)_{F}$ family symmetry, and take $\gamma =\frac{%
\pi }{2}$. This has the effect of reducing the uncertainty in our prediction
Eq.~(\ref{Vcb}) from 0.007 to 0.004. Another prediction for the ratio: 
\begin{equation}
\left| \frac{V_{ub}}{V_{cb}}\right| =\sqrt{\frac{m_{u}}{m_{c}}}
\end{equation}
is quite general for models with ``nearest-neighbour'' mixing \cite{review}.

\subsubsection{Scenario B: ``up quark mass $m_{u}$ walking around the
corner, while down quark mass $m_{d}$ runs along the diagonal''}

Now the mass matrices for the down quarks $M_{D}$ and charged leptons $M_{E}$
are supposed to have the same form as in Eq.~(\ref{LFM1}), while the
Hermitian mass matrix for the up quarks is taken to be: 
\begin{equation}
M_{U}=\pmatrix{ 0 & 0 & c_U \cr 0 & A_U & 0 \cr c_U^{\ast} & 0 & B_ U \cr}
\label{LFM2}
\end{equation}
All the elements of $M_{U}$ can again be readily determined in terms of the
physical masses as: 
\begin{equation}
A_{U}=m_{c}\qquad B_{U}=m_{t}-m_{u}\qquad \left| c_{U}\right| =\sqrt{%
m_{u}m_{t}}
\end{equation}
The quark mass matrices are diagonalised again by unitary transformations as
in Eq.~(\ref{VUVD}), provided that the matrix $V_{U}$ is changed to 
\begin{equation}
V_{U}=R_{13}^{U}\Phi ^{U}  \label{V1}
\end{equation}
where the 1-3 plane rotation of the $u$ and $t$ quarks and the phase matrix $%
\Phi ^{U}$ (depending on the phase of the element $c_{U}=\left| c_{U}\right|
e^{i\alpha _{U}}$) are parameterised in the following way: 
\begin{equation}
R_{13}^{U}=\pmatrix{ c_{13} & 0 & s_{13} \cr 0 & 1 & 0 \cr -s_{13} & 0 &
c_{13} \cr}\qquad \qquad \Phi ^{U}=\pmatrix{ e^{i \alpha_U} & 0 & 0 \cr 0 &
1 & 0 \cr 0 & 0 & 1 \cr}
\end{equation}
Here $s_{13}\equiv \sin \theta _{13}$ and $c_{13}\equiv \cos \theta _{13}$.

The structure of the CKM matrix now differs from that of Eq.~(\ref{CKM1}) as
it contains the direct 1-3 plane rotation for the up quarks: 
\begin{equation}
V=V_{U}V_{D}^{\dagger }=R_{13}^{U}\Phi ^{U}\left( \Phi ^{D}\right)
^{*}\left( R_{23}^{D}\right) ^{-1}\left( R_{12}^{D}\right) ^{-1}
\label{CKM2}
\end{equation}
although the phases and rotations associated with the down quarks are left
the same as before. This natural parameterisation is now quite close to the
standard one \cite{data}. The proper mixing angles and CP violating phase
(after a suitable re-phasing of the c quark, $c\rightarrow ce^{-i\beta _{D}}$%
) are given by the simple and compact formulae: 
\begin{equation}
\left| V_{us}\right| \simeq s_{12}=\sqrt{\frac{m_{d}}{m_{s}}}\qquad \left|
V_{cb}\right| \simeq s_{23}=\sqrt{\frac{m_{d}}{m_{b}}}\qquad \left|
V_{ub}\right| \simeq s_{13}=\sqrt{\frac{m_{u}}{m_{t}}}  \label{angles}
\end{equation}
and 
\begin{equation}
\delta =\alpha _{U}-\alpha _{D}-\beta _{D}
\end{equation}
While the values of $\left| V_{us}\right| $ and $\left| V_{cb}\right| $ are
practically the same as in scenario A and in good agreement with experiment,
a new prediction for $\left| V_{ub}\right| $ (not depending on the value of
the CP violating phase) should allow experiment to differentiate between the
two scenarios in the near future.

\subsubsection{The CKM Matrix}

Our numerical results for both versions of our model, with a maximal CP
violating phase (see discussion in section 3.3.6), are summarized in the
following CKM matrix: 
\begin{equation}
V_{CKM}=\pmatrix{ 0.975(5) & 0.222(4) & 0.0023(5) \ A \cr & & 0.0036(6) \ B
\cr 0.222(4) & 0.975(8) & 0.038(4) \cr 0.0084(18) & 0.038(4) & 0.999(5) \cr}
\label{ckm}
\end{equation}
The uncertainties in brackets are largely given by the uncertainties in the
quark masses. There is clearly a real and testable difference between
scenarios A and B given by the value of the $V_{ub}$ element. Agreement with
the experimental values of the already known CKM matrix elements \cite{data}
looks quite impressive. The distinctive predictions for the presently
relatively poorly known $V_{ub}$ and $V_{td}$ elements\footnote{%
Present data from CLEO \cite{cleo} and LEP \cite{lepvub} favour scenario B.}
should be tested in the near future, by experimental data from the
B-factories.

\subsection{Lepton Sector}

The lepton mixing matrix is defined analogously to the CKM matrix: 
\begin{equation}
U=U_{\nu }U{_{E}}^{\dagger }  \label{U}
\end{equation}
where the indices $\nu $ and $E$ stand for $\nu =(\nu _{e},\nu _{\mu },
\nu_{\tau })$ and $E=(e,\mu ,\tau )$. Our model predicts the charged lepton
mixing angles in the matrix $U_{E}$ with high accuracy to be: 
\begin{equation}
\sin \theta _{e\mu }=\sqrt{\frac{m_{e}}{m_{\mu }}}\qquad \sin \theta _{\mu
\tau }=\sqrt{\frac{m_{e}}{m_{\tau }}}\qquad \sin \theta _{e\tau }\simeq 0
\label{charged}
\end{equation}
provided that the charged lepton masses follow alternative I, along with the
down quarks ones, or

\begin{equation}
\sin \theta _{e\mu }=0\qquad \sin \theta _{\mu \tau }=0\qquad \sin \theta
_{e\tau }=\sqrt{\frac{m_{e}}{m_{\tau }}}  \label{charged'}
\end{equation}
if they follow alternative II, like the up quarks. We consider only
alternative I for the charged leptons in our $SU(3)_F$ family symmetry
model, since the alternative II for them would lead in general to the same
hierarchy between their masses as appears for the up quarks, which is
experimentally excluded.

However, in both cases, these small charged lepton mixing angles will not
markedly effect atmospheric neutrino oscillations \cite{superkam}, which
appear to require essentially maximal mixing $\sin ^{2}2\theta _{atm}\simeq
1 $. It follows then that the large neutrino mixing responsible for
atmospheric neutrino oscillations should mainly come from the $U_{\nu }$
matrix associated with the neutrino mass matrix in (\ref{U}). This requires
a different mass matrix texture for the neutrinos compared to the charged
fermions; in this connection a number of interesting models have been
suggested \cite{hall}, using different mechanisms for neutrino mass
generation. Remarkably, there appears to be no need in our case for some
different mechanism to generate the observed mixing pattern of neutrinos:
they can get physical masses and mixings via the usual ``see-saw'' mechanism 
\cite{GRS}, using the proposed LFMG mechanism for their primary Dirac and
Majorana masses.

Let us consider this pattern in more detail. According to the ``see-saw''
mechanism the effective mass-matrix $M_{\nu}$ for physical neutrinos has the
form \medskip \cite{GRS} 
\begin{equation}
M_{\nu }=-M_{N}^{T}M_{NN}^{-1}M_{N}  \label{nu}
\end{equation}
where $M_{N}$ is their Dirac mass matrix, while $M_{NN}$ is the Majorana
mass matrix of their right-handed components. Therefore, one must first
choose which of the two alternatives I or II works for the neutrino
mass-matrix (\ref{nu}), particularly for its Dirac and Majorana ingredients, 
$M_{N}$ and $M_{NN}$ respectively. There are in fact four possible cases:
(i) alternative I works for both the matrices $M_{N}$ and $M_{NN}$ ; (ii)
alternative I works for $M_{N}$, while alternative II works for $M_{NN}$ ;
(iii) alternative I works for $M_{NN}$ , while alternative II works for $%
M_{N}$ ; (iv) alternative II works for both the matrices $M_{N}$ and $M_{NN}$%
. It is easy to confirm that the cases (iii) and (iv) are completely
excluded, since they do not lead to any significant $\nu _{\mu }-\nu _{\tau
} $ mixing as is required by the SuperKamiokande data \cite{superkam}.
However the first two cases, (i) and (ii), happen to be of actual
experimental interest when an appropriate hierarchy is proposed between the
matrix elements in the matrices $M_{N}$ and $M_{NN}$. So, we also have two
possible scenarios in the lepton sector.

\subsubsection{Scenario A*: ``all the lightest lepton Dirac and Majorana
masses running along the diagonal''}

Let us enlarge first on the case (i) considered recently in ref.~\cite
{Fukuyama}. The eigenvalues of the neutrino Dirac mass matrix $M_{N}$ are
taken to have a hierarchy similar to that for the charged leptons (and down
quarks)

\begin{equation}
M_{N3}:M_{N2}:M_{N1}\simeq 1:y^{2}:y^{4},\quad y\approx 0.1  \label{h1}
\end{equation}
%\medskip 
and the eigenvalues of the Majorana mass matrix $M_{NN}$ are taken to have a
stronger hierarchy

\begin{equation}
M_{NN3}:M_{NN2}:M_{NN1}\simeq 1:y^{4}:y^{6}  \label{h2}
\end{equation}
One then readily determines the general LFMG matrices $M_{N}$ and $M_{NN}$
to be of the type

\begin{equation}
M_{N}\simeq M_{N3}\left( 
\begin{array}{lll}
0 & \alpha y^{3} & 0 \\ 
\alpha y^{3} & y^{2} & \alpha y^{2} \\ 
0 & \alpha y^{2} & 1
\end{array}
\right)  \label{m0}
\end{equation}
and

\begin{equation}
M_{NN}\simeq M_{NN3}\left( 
\begin{array}{lll}
0 & \beta y^{5} & 0 \\ 
\beta y^{5} & y^{4} & \beta y^{3} \\ 
0 & \beta y^{3} & 1
\end{array}
\right)  \label{m00}
\end{equation}
where, for both the order-one parameters $\alpha $ and $\beta $ contained 
in the matrices $M_{N}$ and $M_{NN}$, we take an extra condition of
the type

\begin{equation}
\left| \Delta -1\right| \compoundrel<\over\sim y^{2}\qquad (\Delta \equiv
\alpha ,\beta )  \label{natur}
\end{equation}
according to which they are supposed to be equal to unity with a few percent
accuracy. Substitution in the seesaw formula (\ref{nu}) generates an
effective physical neutrino mass matrix $M_{\nu }$ of the form:

\begin{equation}
M_{\nu }\simeq -\frac{M_{N3}^{2}}{M_{NN3}}\left( 
\begin{array}{lll}
0 & y & 0 \\ 
y & 1+(y-y^{2})^{2} & 1-(y-y^{2}) \\ 
0 & 1-(y-y^{2}) & 1
\end{array}
\right)  \label{matr1}
\end{equation}
%\medskip 
The physical neutrino masses are then given with an accuracy of $O(y)$ taken
by:

\begin{eqnarray}
m_{\nu 1} &\simeq &(\frac{1}{2}-\frac{\sqrt{3}}{2})\frac{M_{N3}^{2}}{M_{NN3}}%
\cdot y, \\
m_{\nu 2} &\simeq &(\frac{1}{2}+\frac{\sqrt{3}}{2})\frac{M_{N3}^{2}}{M_{NN3}}%
\cdot y,\quad m_{\nu 3}\simeq (2-y)\frac{M_{N3}^{2}}{M_{NN3}}  \nonumber
\end{eqnarray}
Taking the Dirac mass parameter $M_{N3}$ to approximately equal the top
quark mass $m_{t}$, the Majorana mass parameter can be determined to be $%
M_{NN3}\simeq 1\cdot 10^{15}$ GeV from the experimentally observed value of $%
\Delta m_{atm}^{2}\simeq 3\cdot 10^{-3}$ eV$^{2}$ \cite{superkam} (one may
neglect the changes in the lepton masses and mixings coming from running
from the top mass scale to the MeV scale).

The mass matrix (\ref{matr1}) gives essentially maximal $\nu _{\mu }-\nu
_{\tau }$ mixing in agreement with the atmospheric neutrino data \cite
{superkam}. Also it is in agreement, while marginal, with the known LMA
(large mixing angle MSW) fit to the solar neutrino experiments \cite
{nu-expSMR}. The standard two flavour atmospheric and solar neutrino
oscillation parameters are expressed in terms of the leptonic CKM matrix $U$
as follows: 
\begin{eqnarray}
\sin ^{2}2\theta _{sun} &=&4\left| U_{e1}\right| ^{2}\left| U_{e2}\right|
^{2},\quad \sin ^{2}2\theta _{atm}=4\left| U_{\mu 3}\right| ^{2}(1-\left|
U_{\mu 3}\right| ^{2})  \label{fac} \\
\Delta m_{sun}^{2} &=&m_{\nu 2}^{2}-m_{\nu 1}^{2},\quad \Delta
m_{atm}^{2}=m_{\nu 3}^{2}-m_{\nu 2}^{2}  \nonumber
\end{eqnarray}
assuming the approximate decoupling condition $/U_{e3}/^{2}\ll 1$ (see
below). The predicted values of these parameters in scenario A* are \cite
{Fukuyama}: 
\begin{equation}
\sin ^{2}2\theta _{atm}\simeq 1,\quad \sin ^{2}2\theta _{sun}\simeq \frac{2}{%
3},\quad U_{e3}\simeq \frac{1}{2\sqrt{2}}y,\quad \frac{\Delta m_{sun}^{2}}{%
\Delta m_{atm}^{2}}\simeq \frac{\sqrt{3}}{4}y^{2}  \label{pred1}
\end{equation}
to be well compared with the experimentally allowed regions \cite
{superkam,nu-expSMR} corresponding to the LMA solution for the solar
neutrino problem

\begin{eqnarray}
0.82 &\leq &\sin ^{2}2\theta _{atm}\leq 1,\quad 0.65\leq \sin ^{2}2\theta
_{sun}\leq 1  \nonumber \\
\left| U_{e3}\right| ^{2} &\leq &0.05,\quad 3\cdot 10^{-3}\leq \frac{\Delta
m_{sun}^{2}}{\Delta m_{atm}^{2}}\leq 5\cdot 10^{-2}  \label{exp}
\end{eqnarray}

\subsubsection{ Scenario B*: ``the lightest Majorana mass walking around the
corner, while the lightest Dirac masses run along the diagonal''}

We now turn to case (ii), which appears to lead to another case of neutrino
oscillations where together with essentially maximal mixing of atmospheric
neutrinos the small mixing angle solution 
MSW (SMA) naturally appears for solar
neutrinos. We again take the alternative I form (\ref{m0}) for the Dirac
mass matrix $M_{N}$, while for the Majorana mass
matrix $M_{NN}$ we take the general alternative II form

\begin{equation}
M_{NN}\simeq M_{NN3}\left( 
\begin{array}{lll}
0 & 0 & \beta y^{q} \\ 
0 & y^{p} & 0 \\ 
\beta y^{q} & 0 & 1
\end{array}
\right)  \label{m000}
\end{equation}
with an arbitrary eigenvalue hierarchy of the type

\begin{equation}
M_{NN3}:M_{NN2}:M_{NN1}\simeq 1:y^{p}:y^{2q}  \label{h3}
\end{equation}
where $p$ and $q$ are positive integers. Again, for the parameters $\alpha $ 
and $\beta $ in the mass-matrices $M_{N}$ and $M_{NN}$, the extra condition 
(\ref{natur}) is assumed. The seesaw formula (\ref{nu}) then generates an
effective physical neutrino mass matrix $M_{\nu }$ which for any $p\geq 2q-1$%
, $q>1$ has a particularly simple form (inessential higher order terms are
omitted)

\begin{equation}
M_{\nu }\simeq -\frac{M_{N3}^{2}}{M_{NN3}}y^{4-p}\left( 
\begin{array}{lll}
y^{2} & y & y \\ 
y & 1-y^{p-2q+2} & 1+y^{p-q-1} \\ 
y & 1+y^{p-q-1} & 1
\end{array}
\right)  \label{matr2}
\end{equation}
One can see that the matrix (\ref{matr2}) automatically leads to the large
(maximal) $\nu _{\mu }-\nu _{\tau }$ mixing and small $\nu _{e}-\nu _{\mu }$
mixing for any hierarchy in the Majorana mass matrix (\ref{m000}) satisfying
the condition $p\geq 2q-1$, $q>1$ mentioned above. Taking, for an example, $%
p=5$ and $q=3$ one can find for neutrino masses the values

\begin{equation}
m_{\nu 1}\simeq \frac{M_{N3}^{2}}{M_{NN3}}\cdot \frac{y^{2}}{3}, \qquad
m_{\nu 2}\simeq -\frac{3}{2}\frac{M_{N3}^{2}}{M_{NN3}}, \qquad m_{\nu
3}\simeq 2\frac{M_{N3}^{2}}{M_{NN3}}\cdot y^{-1}  \label{masses2}
\end{equation}
while the following predictions for the two flavour oscillation parameters
naturally appear

%\medskip
\begin{equation}
\sin ^{2}2\theta _{atm}\simeq 1,\quad \sin ^{2}2\theta _{sun}\simeq \frac{2}{%
9}y^{2},\quad U_{e3}\simeq \frac{1}{\sqrt{2}}y,\quad \frac{\Delta m_{sun}^{2}%
}{\Delta m_{atm}^{2}}\simeq \frac{9}{16}y^{2}  \label{pred2}
\end{equation}
Again taking $y\approx 0.1$, 
%For the known value of the hierarchy parameter $y\approx
%(m_{s}/m_{b})^{1/2}\approx 0.15$ in the model, 
our predictions (\ref{pred2}) turn out to be inside of the experimentally
allowed intervals \cite{nu-expSMR} for the SMA solution for solar neutrino
oscillations:

\begin{eqnarray}
0.82 &\leq &\sin ^{2}2\theta _{atm}\leq 1,\quad 10^{-3}\leq \sin ^{2}2\theta
_{sun}\leq 10^{-2}  \nonumber \\
\left| U_{e3}\right| ^{2} &\leq &0.05,\quad 5\cdot 10^{-4}\leq \frac{\Delta
m_{sun}^{2}}{\Delta m_{atm}^{2}}\leq 9\cdot 10^{-3}  \label{sma}
\end{eqnarray}
Note that in contrast to the LMA case~(\ref{pred1}), one must include now
even the small contributions stemming from the charged lepton sector (see
Eq.~(\ref{charged}) into the solar neutrino oscillations.

So, one can see that the LFMG mechanism works quite successfully in the
lepton sector as well as in the quark sector. Remarkably, the same mechanism
results simultaneously in small quark mixing and large lepton mixing (in
scenario B$^{*}$ for the second and third families). On the other hand,
there could be different sources of the large neutrino mixing. An attractive
alternative method for generating large neutrino mixing within the
supersymmetric Standard Model is via $R$-parity violating interactions,
which can give radiatively induced neutrino masses and mixing angles in just
the area required by the current observational data \cite{chun,lep}.

\section{The SU(3)$_{F}$ Theory of Flavour Mixing}

\subsection{Motivation}

The choice of a local chiral $SU(3)_{F}$ symmetry \cite{jon} as the
underlying flavour symmetry beyond the Standard Model is based on the
following three ``pillars'':

(i) It provides a natural explanation of the number three of observed
quark-lepton families, correlated with three species of massless or light ($%
m_{\nu}<M_{Z}/2$) neutrinos contributing to the invisible Z boson partial
decay width \cite{data};

(ii) Its local nature could be expected by analogy with the other
fundamental symmetries of the Standard Model, such as weak isospin symmetry $%
SU(2)_{W}$ or colour symmetry $SU(3)_{C}$. A stronger motivation follows
from superstrings (at least it is one of the most important
model-independent results derived from perturbative string theory); there
are no global non-abelian symmetries in the 4D string models (for a review
see \cite{Sstring});

(iii) Any family symmetry should be completely broken at low energies in
order to conform with reality. This symmetry should be chiral, rather than a
vectorlike one under which left-handed and right-handed fermions have the
same group-theoretical features. A vectorlike symmetry would not provide any
mass protection and would give degenerate rather than hierarchical quark and
lepton family mass spectra. Interestingly, both known examples of local
vectorlike symmetries, electromagnetic $U(1)_{EM}$ and colour $SU(3)_{C}$,
appear to be exact symmetries. Also, in standard GUT models fermions and
antifermions can lie in the same irreducible representation; so a $%
GUT\otimes SU(3)_H$ model necessarily has a chiral family symmetry.

So, if one takes these naturality criteria seriously, all the candidates for
flavour symmetry can be excluded except for local chiral $SU(3)_{F}$
symmetry. The chiral $U(1)$ symmetries \cite{fn1,leurer,ibanezross} do not
satisfy the criterion (i) and are in fact applicable to any number of
quark-lepton families. Also, the chiral $SU(2)$ symmetry can contain,
besides two light families treated as its doublets, any number of additional
(singlets or new doublets of $SU(2)$) families. All the global non-abelian
symmetries are excluded by criterion (ii), while the vectorlike ones are
excluded by the last criterion (iii).

Among the applications of the $SU(3)_{F}$ symmetry, the most interesting
ones are the description of the quark and lepton masses and mixings in the
Standard Model and GUTs \cite{jon}, neutrino masses and oscillations \cite
{ber1} and rare processes \cite{jon0}. Remarkably, the $SU(3)_{F}$ invariant
Yukawa coupling are always accompanied by an accidental global chiral $U(1)$
symmetry \cite{jon0,q84}, which can be identified with the Peccei-Quinn
symmetry \cite{peccei} provided it is not explicitly broken in the Higgs
sector, thus giving a solution to the strong CP problem. In the SUSY context 
\cite{q86}, the $SU(3)_{F}$ model leads to a special relation between
fermion masses and the soft SUSY breaking terms at the GUT scale, so that
all the dangerous supersymmetric flavour-changing processes are naturally
suppressed \cite{ber2}.

Another sector of applications is related with a new type of topological
defect---non-abelian cosmic strings appearing during the spontaneous
breaking of the $SU(3)_{F}$---considered as a possible candidate for the
cold dark matter in the Universe \cite{jon1}. And the last point worthy of
note is that the local chiral $SU(3)_{F}$ symmetry has been applied to GUTs
not only in a direct product form, such as $SU(5)\otimes SU(3)_{F}$ \cite
{jon}, $SO(10)\otimes SU(3)_{F}$ \cite{wu} or $E(6)\otimes SU(3)_{F}$ \cite
{wu}, but also as a subgroup of the family unified $SU(8)$ GUT model \cite
{jon2}.

%\medskip

\subsection{The Matter and Higgs Supermultiplets}

In the MSSM extended by the local chiral $SU(3)_{F}$ symmetry the quark and
lepton superfields are supposed to be $SU(3)_{F}$ chiral triplets, so that
their left-handed (weak-doublet) components

\begin{eqnarray}
L_{\alpha f} &=&\left[ \left( 
\begin{array}{c}
U_{L} \\ 
D_{L}
\end{array}
\right) _{\alpha }, \, \left( 
\begin{array}{c}
N_{L} \\ 
E_{L}
\end{array}
\right) _{\alpha }\right] =  \label{left} \\
&&  \nonumber \\
&=&\left[ \left( \left( 
\begin{array}{c}
u \\ 
d
\end{array}
\right) _{L},\left( 
\begin{array}{c}
c \\ 
s
\end{array}
\right) _{L},\left( 
\begin{array}{c}
t \\ 
b
\end{array}
\right) _{L}\right), \, \left( \left( 
\begin{array}{c}
\nu _{e} \\ 
e
\end{array}
\right) _{L},\left( 
\begin{array}{c}
\nu _{\mu } \\ 
\mu
\end{array}
\right) _{L},\left( 
\begin{array}{c}
\nu _{\tau } \\ 
\tau
\end{array}
\right) _{L}\right) \right]  \nonumber
\end{eqnarray}
are triplets of $SU(3)_{F}$ , while their right-handed (weak-singlet)
components 
\begin{eqnarray}
R_{uf}^{\alpha } &=&[U_{R}^{\alpha } \, ,N_{R}^{\alpha}] =[(u,c,t)_{R}\, ,
(N_{e},N_{\mu },N_{\tau })_{R}]  \label{right} \\
R_{df}^{\alpha } &=&[D_{R}^{\alpha } \, ,E_{R}^{\alpha}] =[(d,s,b)_{R} \,
,(e,\mu ,\tau )_{R}]  \nonumber
\end{eqnarray}
are anti-triplets (or vice versa). For completeness we have included the
right-handed neutrino superfields $N_{R}^{\alpha }$ as well. Here $\alpha $
is a family symmetry index ($\alpha =1,2,3)$, while the index $f$ simply
stands for the type of basic fermion, quark ($f=q$) or lepton ($f=l$).

The matter sector also includes some set of right-handed states, which are
needed to cancel all the $SU(3)_{F}$ triangle anomalies. While being
singlets under the Standard Model $SU(3)_{C}\otimes SU(2)_{W}\otimes
U(1)_{Y} $ gauge symmetry, these states can be chosen as sixteen
right-handed triplets $r_{\alpha }^{n}$ ($n=1,..,16$) or some other
combination of $SU(3)_{F}$ multiplets properly compensating the triangle
anomalies. All of them receive heavy masses of order the family symmetry
breaking scale $M_{F}$ and are practically decoupled from the low-energy
particle spectrum. In fact, they look like right-handed heavy neutrinos and,
as such, can contribute to the ordinary light neutrino masses and mixings
via the ``see-saw'' mechanism \cite{GRS} (see below).

In addition to the standard weak-doublet scalar supermultiplets $H$ and $%
\overline{H}$ of the MSSM, the Higgs sector contains the $SU(3)_{F}$
symmetry-breaking horizontal scalars. These scalars are anti-triplets $%
\overline{\eta} ^{i}$ and sextets $\chi ^{j}$ of $SU(3)_{F}$ (indices $i$
and $j$ number the scalar multiplets, $i=1,2,...$, $j=1,2,...$), which
transform under $SU(3)_{F}$ like matter bi-linears so as to give fermion
masses through the symmetry-allowed general effective Yukawa couplings:

\begin{eqnarray}
W_{Y} &=&\overline{L}_{f}^{\alpha }R_{uf}^{\beta }H_{u}[A_{uf}^{i}\cdot 
\frac{\chi _{\{\alpha \beta \}}^{i}}{M_{F}}+B_{uf}^{i}\cdot \frac{\overline{%
\eta }_{[\alpha \beta ]}^{i}}{M_{F}}]+  \label{yuk} \\
&&+\overline{L}_{f}^{\alpha }R_{df}^{\beta }\overline{H}_{d}[A_{df}^{i}\cdot 
\frac{\chi _{\{\alpha \beta \}}^{i}}{M_{F}}+B_{df}^{i}\cdot \frac{\overline{%
\eta }_{[\alpha \beta ]}^{i}}{M_{F}}]+  \nonumber \\
&&+N_{R}^{\alpha }N_{R}^{\beta }[A_{NN}^{i}\cdot \chi _{\{\alpha \beta
\}}^{i}]  \nonumber
\end{eqnarray}
for the up and down fermions, quarks ($f=q$) and leptons ($f=l$). Also
Majorana couplings are included for the right-handed neutrinos, where only
the symmetrical parts are allowed to appear. The constants $A_{uf,df,N}^{i}$
and $B_{uf,df,N}^{i}$ ($A_{uq}^{i}\equiv A_{U}^{i},$ $A_{ul}^{i}\equiv
A_{N}^{i}$ and so on) are the dimensionless Yukawa constants associated with
the symmetric and anti-symmetric couplings in family space respectively (we
have here used the anti-symmetric feature of triplets in $SU(3)$: $\overline{%
\eta }_{[\alpha \beta ]}\equiv \epsilon _{\alpha \beta \gamma }\overline{%
\eta }^{\gamma }$). These couplings normally appear via the well-known
see-saw type mechanism \cite{fn1,jon} due to the exchange of a special set
of heavy (of order the flavour scale $M_{F}$) vectorlike fermions or
directly through the gravitational interactions, being suppressed in the
latter case by inverse powers of the Planck scale $M_{Pl}$ \cite{jon3}. One
can even make the Yukawa couplings (\ref{yuk}) renormalizable, by
introducing additional scalars with both electroweak and horizontal indices.
The VEVs of the horizontal scalars $\overline{\eta }^{i}$ and $\chi ^{i}$,
taken in general as large as $M_{F}$ (or $M_{Pl}$), are supposed to be
hierarchically arranged along the different directions in family space, so
as to properly imitate the observed quark and lepton mass and mixing
hierarchies even with coupling constants $A_{uf,df,N}^{i}$ and $%
B_{uf,df,N}^{i}$ taken of the order $O(1)$. The Majorana couplings of the
right-handed neutrinos $N_{R}^{\alpha }$ do not, of course, involve the
Weinberg-Salam Higgs field $H_{u}$ and are thus naturally of order $O(M_{F})$%
. As was mentioned above, the Yukawa couplings (\ref{yuk}) admit an extra
global chiral symmetry $U(1)$ identified with the Peccei-Quinn symmetry%
\footnote{%
Even in the non-supersymmetric Standard Model which contains only one Higgs
doublet this symmetry can be successfully implemented \cite{jon,q84}, by
assigning the $U(1)_{PQ}$ charges directly to the horizontal scalars $\eta
^{i}$ and $\chi ^{i}$.} $U(1)_{PQ}$ \cite{peccei}. However, in general, one
must also consider the Higgs superpotential (see below), in order to
determine whether there exists global $U(1)$ or discrete $Z_{N}$
symmetry(ies) in the model.

There is a more economic way of generating effective Yukawa couplings, by
using just horizontal triplets $\eta _{\alpha }^{i}$ and anti-triplets $%
\overline{\eta }_{[\alpha \beta ]}^{i}$ and imitating sextets by the pairing
of triplet scalars. Instead of the couplings (\ref{yuk}), in this case \cite
{jon0} one has: 
\begin{eqnarray}
W_{Y} &=&\overline{L}_{f}^{\alpha }R_{uf}^{\beta }H_{u}[A_{uf}^{ij}\cdot %
\mbox{\boldmath$ \eta$}_{\alpha }^{i}\mbox{\boldmath$ \eta$}_{\beta }^{j}%
{\bf I}^{k_{uf}^{ij}}+B_{uf}^{i}\cdot \overline{\mbox{\boldmath$ \eta$}}%
_{[\alpha \beta ]}^{i}{\bf I}^{n_{uf}^{i}}]+  \label{yuk'} \\
&&+\overline{L}_{f}^{\alpha }R_{df}^{\beta }\overline{H}_{d}[A_{df}^{ij}%
\cdot \mbox{\boldmath$ \eta$}_{\alpha }^{i}\mbox{\boldmath$ \eta$}_{\beta
}^{j}{\bf I}^{k_{df}^{ij}}+B_{df}^{i}\cdot \overline{\mbox{\boldmath$  \eta$}%
}_{[\alpha \beta ]}^{i}{\bf I}^{n_{df}^{i}}]+  \nonumber \\
&&+N_{R}^{\alpha }N_{R}^{\beta }[A_{NN}^{ij}\cdot \mbox{\boldmath$ \eta $}%
_{\alpha }^{i}\mbox{\boldmath$ \eta$}_{\beta }^{j}{\bf I}^{k_{NN}^{ij}}] 
\nonumber
\end{eqnarray}
where all the horizontal ``bold'' scalar fields ${\bf \eta }^{i}$ are taken
now to be scaled by the $SU(3)_{F}$ symmetry breaking scale $M_{F}$, $%
\mbox{\boldmath$ \eta $}_{\alpha }^{i}\equiv \eta _{\alpha }^{i}/M_{F}$. The
hierarchy in the corresponding mass matrices (including the up-down
hierarchy) depends solely on the powers $k_{u,d,N}^{ij}$ and $n_{u,d,N}^{i}$
of the specially introduced hierarchy-making singlet scalar field ${\bf I}$ (%
${\bf I\equiv }I{\bf /}M_{F}$) in (\ref{yuk'}). This description turns out
to be well fitted to the supersymmetric case where both types of scalar, $%
\eta _{\alpha }^{i}$ and its SUSY counter-part $\overline{\eta }^{i\alpha
}\equiv \epsilon ^{\alpha \beta \gamma }\overline{\eta }_{[\beta \gamma
]}^{i}$, appear automatically.

The basic Yukawa couplings (\ref{yuk}) (or (\ref{yuk'})) can lead to various
types of fermion mass matrices, depending on the vacuum configurations
developed by the horizontal triplets $\eta ^{i}$, anti-triplets $\overline{%
\eta }^{i}$ and sextets $\chi ^{i}$ involved. Remarkably, a texture-zero
structure emerges as one of the generic features of a typical $SU(3)_{F}$
symmetry breaking pattern. This is controlled by the antisymmetric
(anti-triplet) VEVs $<\overline{\eta }^{i}>$ giving off-diagonal elements in
the mass matrices, while the symmetric (sextet) ones $<\chi ^{i}>$ generate
the dominant diagonal elements. Therefore, any of the texture-zero models
catalogued in the literature (see \cite{rrr}) are readily available in the
framework of the $SU(3)_{F}$ model.

For example, the original Fritzsch ansatz \cite{fritzsch2} with the zero
mass-matrix elements, $M_{11}=M_{22}=M_{13}=0$, while presently excluded by
experiment, appears from a basic vacuum configuration of the Higgs potential 
\cite{jon} (or superpotential \cite{q86}) with a simple set of horizontal
scalars---two anti-triplets $\overline{\eta} _{[\alpha \beta ]}$ and $%
\overline{\xi}_{[\alpha \beta ]}$ and one sextet $\chi _{\{\alpha \beta \}}$
developing VEVs along the following directions in family space\footnote{%
Interestingly, the VEVs $\chi _{33}$ and $\overline{\eta} _{[12]}$ in the
coherent basis correspond to the flavour democracy picture \cite{review} for
the mass-matrices of quarks and leptons with symmetrical and antisymmetrical
``pairing forces'' respectively, while the VEV $\xi _{[23]}$ breaks this
picture appropriately.}:

\begin{equation}
\overline{\eta} _{[12]}=-\overline{\eta}_{[21]},\quad \overline{\xi}
_{[23]}=-\overline{\xi}_{[32]},\quad \chi_{\{33\}}  \label{vevs}
\end{equation}
The ``improved'' Fritzsch ansatz (see \cite{review}) with a non-zero $M_{22}$
element appears, when a second sextet $\omega _{\{\alpha \beta \}}$ is
included\footnote{%
Both of these ans\"{a}tze, original and ``improved'', readily appear in a
pure triplet realization (\ref{yuk'}) as well, with three and four scalar
triplets respectively.}, which develops a VEV $\omega _{\{22\}}$.
Unfortunately, this ansatz is not completely predictive of the CKM matrix in
terms of quark mass ratios (which would require no more than three
independent parameters for each of the mass matrices). In particular it
gives no possibility of predicting the $V_{cb}$ element in the CKM matrix.

As a matter of fact the only ansatz left, which successfully predicts all
the CKM matrix elements in terms of quark mass ratios (not to mention here
some eclectic approaches using ad hoc GUT and string arguments, which are
already largely excluded by experiment) is that suggested by the LFMG
mechanism. In the next subsection, we show how the LFMG can be derived
within the $SU(3)_{F}$ model.

\subsection{Minimal flavour mixing from family symmetry}

\subsubsection{Basic vacuum configurations}

We start with an essential question: how does one get the proportionality
condition (\ref{prop}) underlying the LFMG in the framework of the $%
SU(3)_{F} $ model? In terms of the Yukawa superpotential (\ref{yuk}), this
condition (\ref{prop}) means that not only the VEVs of the horizontal
supermultiplets, sextets $\chi _{\{\alpha \beta \}}^{i}$ and anti-triplets $%
\overline{\eta} _{[\alpha \beta ]}^{i}$ of $SU(3)_{F}$ , but also their
Yukawa coupling constants $A_{u,d,N}^{i}$ and $B_{u,d,N}^{i}$ must be
properly arranged. Since the coupling constants are in general quite
independent even in the framework of $SU(3)_{F}$, this in turn means that
the number of horizontal supermultiplets involved in the Yukawa couplings
should be restricted. At the same time, some of these scalars have to
develop VEVs along several directions in the family space in order to
satisfy the proportionality condition (\ref{prop}).

We take the set of scalar supermultiplets in the model to consist of three
pairs of triplets and anti-triplets and two pairs of sextets and
anti-sextets of $SU(3)_{F}$:

\begin{eqnarray}
&&\eta _{\alpha }(\overline{\eta }^{\alpha }),\quad \xi _{\alpha }(\overline{%
\xi }^{\alpha }),\quad \zeta _{\alpha }(\overline{\zeta }^{\alpha }); 
\nonumber  \label{set} \\
&&\chi _{\{\alpha \beta \}}(\overline{\chi }^{_{\{\alpha \beta \}}}),\quad
\omega _{\{\alpha \beta \}}(\overline{\omega }^{_{\{\alpha \beta \}}})
\end{eqnarray}
In section 3.3.2 we will see that just the triplet scalars, to a large 
extent, determine the masses and mixings of the 
quarks and leptons, while the sextets govern the basic vacuum configuration
in family space leading to the LFMG picture for flavour mixing.

We take the Higgs superpotential to have the form

\begin{eqnarray}
W &=&M_{\eta }\eta \overline{\eta }+M_{\xi }\xi \overline{\xi }+M_{\zeta
}\zeta \overline{\zeta }+M_{\chi }\chi \overline{\chi }+M_{\omega }\omega 
\overline{\omega }+  \nonumber \\
&&+f\cdot \epsilon ^{\alpha \beta \gamma }\eta _{\alpha }\xi _{\beta }\zeta
_{_{\gamma }}+\overline{f}\cdot \epsilon _{\alpha \beta \gamma }\overline{%
\eta }^{\alpha }\overline{\xi }^{\beta }\overline{\zeta }^{_{\gamma }}
\label{spot} \\
&&+F\cdot \epsilon ^{\alpha \beta \gamma }\epsilon ^{\alpha ^{\prime }\beta
^{\prime }\gamma ^{\prime }}\chi _{\{\alpha \alpha ^{\prime }\}}\chi
_{\{\beta \beta ^{\prime }\}}\omega _{\{\gamma \gamma ^{\prime }\}}+%
\overline{F}\cdot \epsilon _{\alpha \beta \gamma }\epsilon _{\alpha ^{\prime
}\beta ^{\prime }\gamma ^{\prime }}\overline{\chi }^{\{\alpha \alpha
^{\prime }\}}\overline{\chi }^{\{\beta \beta ^{\prime }\}}\overline{\omega }%
^{\{\gamma \gamma ^{\prime }\}}  \nonumber
\end{eqnarray}
with masses $M_{\eta ,\xi ,\zeta }$ and $M_{\chi ,\omega }$ and coupling
constants $f$, $\overline{f}$, $F$ and $\overline{F}$. One can readily
notice that all the trilinear couplings of triplets and sextets in the
superpotential $W$ are very specific just to the $SU(3)$ symmetry. These
couplings by themselves induce, as we will see later, strictly orthogonal
(to each other) VEVs for the triplets and sextets involved. However, as was
argued above, some of these scalars also have to develop hierarchically
small parallel VEVs in the family space, so as to get the proportionality
condition (\ref{prop}) satisfied. For this purpose we include in the
superpotential $W$ the additional terms

\begin{equation}
\Delta W_{1}=a\cdot \left( \frac{S}{M_{F}}\right) ^{r}M_{F}\overline{\eta }%
\zeta  \label{spot'}
\end{equation}
with a particular coupling of triplets $\overline{\eta }$ and $\zeta $
which, through their mixing, will cause some extra small parallel VEVs for
them whose magnitudes will be determined by the power $r$ of the VEV of the
singlet scalar $S$. This hierarchy-making scalar $S$ also appears in the
Yukawa couplings (see below). We emphasize here that, besides the $SU(3)_{F}$
symmetry, the superpotential $W$ has a global $U(1)$ symmetry. This global $%
U(1)$ is in substance the custodial symmetry strictly protecting the form of
the superpotential $W$ (\ref{spot}) considered\footnote{%
Let us note that this $U(1)$ may be identified with a superstring-inherited
anomalous $U(1)_{A}$ gauge symmetry \cite{gr-sch} broken at a high scale
through the Fayet-Iliopoulos (FI) $D$-term \cite{witten}. One can take just
this scale as the family symmetry scale $M_{F}$. While for the heterotic
strings this scale normally lies only a few orders of magnitude below the
Planck mass $M_{Pl}$, in the orientifold case, where the FI term appears
moduli-dependent \cite{nilles'}, it can be made to locate much lower.}.

Let us come now to a general analysis of the non-trivial supersymmetric
vacuum configurations of the superpotential $W$. One can quickly find from
the vanishing F-terms of the triplet supermultiplets involved,

\begin{eqnarray}
M_{\eta }\eta _{\alpha }+\overline{f}\cdot \epsilon _{\alpha \beta \gamma }%
\overline{\xi }^{\beta }\overline{\zeta }^{_{\gamma }}+a\cdot \left( \frac{S%
}{M_{F}}\right) ^{r}M_{F}\zeta _{\alpha } &=&0,\quad M_{\eta }\overline{\eta 
}^{\alpha }+f\cdot \epsilon ^{\alpha \beta \gamma }\xi _{\beta }\zeta
_{_{\gamma }}=0,  \nonumber \\
M_{\xi }\xi _{\beta }+\overline{f}\cdot \epsilon _{\alpha \beta \gamma }%
\overline{\eta }^{\alpha }\overline{\zeta }^{_{\gamma }} &=&0,\quad M_{\xi }%
\overline{\xi }^{\beta }+f\cdot \epsilon ^{\alpha \beta \gamma }\eta
_{\alpha }\zeta _{_{\gamma }}=0,  \nonumber  \label{ftrms} \\
M_{\zeta }\zeta _{\gamma }+\overline{f}\cdot \epsilon _{\alpha \beta \gamma }%
\overline{\eta }^{\alpha }\overline{\xi }^{\beta } &=&0,  \nonumber \\
M_{\zeta }\overline{\zeta }^{\gamma }+f\cdot \epsilon ^{\alpha \beta \gamma
}\eta _{\alpha }\xi _{\beta }+a\cdot \left( \frac{S}{M_{F}}\right) ^{r}M_{F}%
\overline{\eta }^{\gamma } &=&0,
\end{eqnarray}
that they develop VEVs satisfying the equations %\medskip

\begin{eqnarray}
\eta _{\alpha }\overline{\eta }^{\alpha } &=&\frac{M_{\xi }M_{\zeta }} {f%
\overline{f}}, \quad \xi _{\alpha }\overline{\xi }^{\alpha }= \frac{M_{\eta
}M_{\zeta }}{f\overline{f}}, \quad \zeta _{\alpha } \overline{\zeta }%
^{\alpha }=\frac{M_{\eta }M_{\xi }}{f\overline{f}}  \label{tvevs} \\
\epsilon ^{\alpha \beta \gamma }\eta _{\alpha }\xi _{\beta }\zeta _{_{\gamma
}} &=&-\frac{M_{\eta }M_{\xi }M_{\zeta }}{f\overline{f}^{2}}, \quad \epsilon
_{\alpha \beta \gamma }\overline{\eta }^{\alpha } \overline{\xi }^{\beta }%
\overline{\zeta }^{_{\gamma }} =-\frac{M_{\eta }M_{\xi }M_{\zeta }}{%
\overline{f}^{2}f}  \nonumber
\end{eqnarray}
All but one of the scalar products between pairs of triplets vanish

\begin{eqnarray}
\eta _{\alpha }\overline{\xi }^{\alpha } &=&\xi _{\alpha } \overline{\eta }%
^{\alpha }=\xi _{\alpha }\overline{\zeta }^{\alpha } =\zeta _{\alpha } 
\overline{\eta }^{\alpha } =\zeta _{\alpha }\overline{\xi }^{\alpha }=0
\label{ort} \\
\eta _{\alpha }\overline{\zeta }^{\alpha } &=&-\frac{a}{f\overline{f}}\cdot
\left( \frac{S}{M_{F}}\right)^rM_FM_{\xi }  \nonumber
\end{eqnarray}
and their non-zero components can be chosen as follows:

\begin{equation}
\begin{array}{ll}
\eta _{\alpha }=(X,0,0), & \overline{\eta }^{\alpha }=(\overline{X},0,%
\overline{x}), \\ 
\xi _{\alpha }=(0,Y,0), & \overline{\xi }^{\alpha }=(0,\overline{Y},0), \\ 
\zeta _{\alpha }=(z,0,Z), & \overline{\zeta }^{\alpha }=(\overline{z},0,%
\overline{Z}).
\end{array}
\label{comp}
\end{equation}
The components $\overline{x}$, $z$ and $\overline{z}$ which appear in them
are small and just arise due to the $\overline{\eta }\zeta $ mixing term in
the superpotential $W$ mentioned above. We propose here that the VEV of the
singlet scalar $S$ responsible for this mixing is typically somewhat smaller
than the basic VEVs of the triplets (\ref{tvevs}), so as to have some input
hierarchy parameter $\frac{S}{M_{F}}$ in the model (see below).

The vanishing of the D-terms

\begin{equation}
(\Phi ,T^{A}\Phi )=0,  \label{d}
\end{equation}
where all the scalars fields (triplets and sextets) are grouped in the
vector $\Phi $ ($T^{A}$ are generators of $SU(3)_{F}$ acting on the
reducible representation given by the vector $\Phi $), results in further
limitations on the possible supersymmetric vacuum configurations.
Particularly, for the generator $T^{4+i5}$, equation (\ref{d}) leads to an
additional non-trivial condition on the VEVs of the triplets $\eta _{\alpha
}(\overline{\eta }^{\alpha })$ and $\zeta _{\alpha }(\overline{\zeta }%
^{\alpha})$ 
\begin{equation}
-\overline{X}^{*}\overline{x}+Z^{*}z-\overline{Z}^{*}\overline{z}=0,
\label{con}
\end{equation}
When this condition is combined with the basic F-term equations (\ref{ftrms}%
), the values of the small VEV components $\overline{x}$, $z$ and $\overline{%
z}$ can be expressed in terms of the VEV of the scalar singlet $S$ (the bold
letter ${\bf S}$ stands for $S{\bf /}M_{F}$) and the ratios of the big VEV
components as follows:

\begin{eqnarray}
\overline{x} &=&-a{\bf S}^{r}M_{F}\frac{\overline{X}}{M_{\zeta }}\frac{Y}{%
M_{\eta }}\frac{\overline{Z}^{*}}{Z^{*}}\frac{1}{1+\mid f\frac{Y}{M_{\eta }}%
\mid ^{2}},  \nonumber \\
z &=&-a{\bf S}^{r}M_{F}\frac{\overline{X}}{M_{\zeta }}\frac{\overline{Z}^{*}%
}{Z^{*}}\frac{1}{1+\mid f\frac{Y}{M_{\eta }}\mid ^{2}},  \label{small} \\
\overline{z} &=&-a{\bf S}^{r}M_{F}\frac{\overline{X}}{M_{\zeta }}  \nonumber
\end{eqnarray}
The VEVs slide by themselves into the valleys determined by the F- and
D-term conditions (\ref{ftrms},\ref{d}) conserving supersymmetry.

Let us turn now to the sextet superfields in the superpotential $W$ (\ref
{spot}). From their vanishing F-term equations

\begin{eqnarray}
M_{\chi }\chi _{\{\alpha \beta \}}+\overline{F}\epsilon _{\alpha \gamma
\sigma }\epsilon _{\beta \delta \rho }\overline{\omega }^{\{\gamma \delta \}}%
\overline{\chi }^{\{\sigma \rho \}} &=&0,M_{\chi }\overline{\chi }^{\{\alpha
\beta \}}+F\epsilon ^{\alpha \gamma \sigma }\epsilon ^{\beta \delta \rho
}\omega _{\{\gamma \delta \}}\chi _{\{\sigma \rho \}}=0,  \nonumber \\
M_{\omega }\omega _{\{\alpha \beta \}}+\overline{F}\epsilon _{\alpha \gamma
\sigma }\epsilon _{\beta \delta \rho }\overline{\chi }^{\{\gamma \delta \}}%
\overline{\chi }^{\{\sigma \rho \}} &=&0,M_{\chi }\overline{\omega }%
^{\{\alpha \beta \}}+F\epsilon ^{\alpha \gamma \sigma }\epsilon ^{\beta
\delta \rho }\chi _{\{\gamma \delta \}}\chi _{\{\sigma \rho \}}=0,  \nonumber
\\
&&  \label{sex}
\end{eqnarray}
and the corresponding D-term equation (\ref{d}), one can confirm that the
sextets develop strictly orthogonal vacuum configurations of the type:

\begin{eqnarray}
\chi _{\{\alpha \beta \}} &=&\chi \cdot diag(1,1,0)_{\alpha \beta },\quad 
\overline{\chi }^{\{\alpha \beta \}}=\overline{\chi }\cdot
diag(1,1,0)^{\alpha \beta }  \label{svevs} \\
\omega _{\{\alpha \beta \}} &=&\omega \cdot diag(0,0,1)_{\alpha \beta
},\quad \overline{\omega }^{\{\alpha \beta \}}=\overline{\omega }\cdot
diag(0,0,1)^{\alpha \beta }  \nonumber
\end{eqnarray}
In order to show this explicitly, one must first rotate away all the
components of the sextet $\chi _{\{\alpha \beta \}}$, except for $\chi _{11}$
and $\chi _{22}$, by the appropriate $SU(3)_{F}$ transformations. Then,
successively using the F-term equations (\ref{sex}), one can see that the
sextet $\omega _{\{\alpha \beta \}}$ and the anti-sextet $\overline{\omega }%
^{\{\alpha \beta \}}$ inescapably develop the VEVs given in equation (\ref
{svevs}) with

\begin{equation}
\omega \overline{\omega }=\frac{M_{\chi }^{2}}{\overline{F}F}  \label{val'}
\end{equation}
while the anti-sextet $\overline{\chi }^{\{\alpha \beta \}}$ also develops
VEVs on the same components, (11) and (22), as the sextet $\chi _{\{\alpha
\beta \}}$ does. Furthermore, a relation appears between them of the form

\begin{equation}
\chi _{11}\overline{\chi }^{11}=\chi _{22}\overline{\chi }^{22}=\frac{%
M_{\chi }M_{\omega }}{\overline{F}F}\quad  \label{val}
\end{equation}
Finally, using the D-term equation (\ref{d}) with generators $T^{3}$ and $%
T^{8}$ and assuming that the $\chi_{\alpha\beta}$ and $\overline{\chi}%
^{\alpha\beta} $ VEVs are large compared with all the other scalar VEVs
contributing to (\ref{d}), one unavoidably comes to the equalities

\begin{equation}
|\chi _{11}|=|\overline{\chi }^{11}|=|\chi _{22}| =|\overline{\chi}^{22}|
\label{vall}
\end{equation}

Thus we are led to the VEVs for the scalars $\chi_{\alpha\beta} $ and $%
\overline{\chi}^{\alpha\beta}$ of the form given in equation (\ref{svevs})
with $|\chi|=|\overline{\chi}|$. These VEVs by themselves break the family $%
SU(3)_{F}$ symmetry to the plane $SO(2)_{F}$ symmetry, acting in the family
subspace of the first and second families of quarks and leptons. 
%This is in
%fact the most symmetrical vacuum configuration caused by the VEVs of sextets 
%$\chi $($\overline{\chi }$) and $\omega $($\overline{\omega }$). 
We will see later that just this vacuum configuration turns out to be
essential for generating Yukawa couplings, which follow our minimal mixing
pattern for quarks and leptons. As mentioned above, we assume the VEVs of
the sextets $\chi_{\alpha\beta} $ and $\overline{\chi}^{\alpha\beta}$ are
the largest in the model, so that this vacuum configuration of the sextets
is not disturbed by the VEVs of the triplets. However, it should be noted
that this assumption in no way influences the hierarchy in masses of the
quarks and leptons, which is completely determined by the triplet scalar
VEVs given above. The relative alignment assigned to the VEVs of the triplet
(\ref{comp}) and the sextet (\ref{svevs}) fields is arranged by introducing
a coupling between the triplet and sextet fields in the superpotential W of
the form

\begin{equation}
\Delta W_{2}=b\cdot \omega \overline{\zeta }\overline{\zeta }\left( \frac{S}{%
M_{F}}\right) ^{p}  \label{acc}
\end{equation}
so as not to disturb the above vacuum configurations. This term also
excludes an accidental global symmetry $U(3)_{triplets}\otimes $ $%
U(3)_{sextets}$ in the superpotential $W$ which might, otherwise, induce the
appearance of extra goldstones (familons) after symmetry breaking. Finally,
one has the total symmetry $SU(3)_{F}\otimes U(1)$ for the superpotential $%
W+\Delta W_{1}+\Delta W_{2}$ considered.

\subsubsection{Yukawa couplings}

One can write down a plethora of $SU(3)_{F}$ invariant Yukawa couplings.
However, by the use of the additional protecting symmetry $U(1)$, one can
choose them in a special form which leads to (at least) one of the four
experimentally allowed scenarios for the minimal flavour mixing of quarks
and leptons. These four scenarios consist of the following four pairs of
combined quark and lepton flavour mixings (see section 2): A+A$^{*}$, A+B$%
^{*}$, B+A$^{*}$ and/or B+B$^{*}$.

Since we have all the triplet superfields needed in our chosen set (\ref{set}%
), the most economic choice of Yukawa couplings would be the pure triplet
realization of the form (\ref{yuk'}) discussed in section 3.2. However it
does not lead to the desired strict proportionality condition (\ref{prop})
or (\ref{ABab}) between mass-matrix elements of the basic (alternative I)
LFMG. So the pure triplet Yukawa couplings can only be used for the up
quarks in scenario B (with its direct 1-3 mixing) for quark mixing. The up
quark Yukawa couplings then take the form\footnote{%
We do not consider here the mass matrix of right-handed neutrinos to which
the scenario B+B$^{*}$ is applicable as well. This will be done later in
section 3.3.5.}:

\begin{equation}
W_{Y1}=\overline{U}_{L}^{\alpha }U_{R}^{\beta }H_{u}[A_{U1}\cdot %
\mbox{\boldmath$ \zeta$}_{\alpha }\mbox{\boldmath $\zeta$}_{\beta
}+A_{U2}\cdot \mbox{\boldmath $\xi$}_{\alpha }\mbox{\boldmath $\xi$}_{\beta }%
{\bf I}^{k}+B_{U}\cdot \overline{\mbox{\boldmath $\xi$}}_{[\alpha \beta ]}%
{\bf I}^{n}]  \label{yukkk}
\end{equation}
which contains two symmetrical couplings related with the bi-linears for the
triplet scalars ${\mbox{\boldmath $\zeta$}}_{\alpha }$ and ${%
\mbox{\boldmath
$\xi$}}_{\alpha }$ and one antisymmetric coupling related with the
anti-triplet $\overline{\mbox{\boldmath $\xi$}}_{[\alpha \beta ]}$ from our
scalar set (\ref{set}). We recall here that the ``bold'' horizontal scalar
fields ${\mbox{\boldmath $\zeta$}}$, \mbox{\boldmath $\xi$}, $\overline{%
\mbox{\boldmath $\xi$}}$ and ${\bf I}$ are scaled by the $SU(3)_{F}$
symmetry mass scale parameter $M_{F}$ (which can generally be as large as
the Plank mass $M_{Pl}$), so that the coupling constants $A_{U,NN}$ and $%
B_{U,NN}$ are all dimensionless and assumed to be of order unity. The
hierarchy in the corresponding mass matrix of the up quarks depends, as in
the general case (\ref{yuk'}), solely on the powers $k$ and $n$ of the
hierarchy-making singlet scalar field ${\bf I}$ whose values are determined
by the $U(1)$ symmetry imposed. According to the triplet VEVs (\ref{comp}),
the first symmetrical coupling in $W_{Y1}$ giving masses to the heaviest
third family are not suppressed, while all other couplings (including the
direct 1-3 mixing term) are suppressed by powers of the scalar ${\bf I}$. As
in the general case (\ref{yuk'}), by introducing additional scalars with
electroweak and horizontal indices combined one could make the Yukawa
couplings (\ref{yukkk}) renormalizable.

The other possible Yukawa couplings, which can be written with the triplets (%
\ref{set}), are those having the form

\begin{eqnarray}
W_{Y2} &=&\overline{L}_{f}^{\alpha }R_{uf}^{\beta }H_{u}[A_{uf}\cdot 
\overline{\mbox{\boldmath $\eta$}}_{[\alpha \sigma ]}\overline{%
\mbox{\boldmath $\eta$}}_{[\beta \rho ]}+B_{uf}\cdot \epsilon _{\alpha \beta
\sigma }\mbox{\boldmath $\eta$}_{\rho }{\bf S}^{l}]\overline{%
\mbox{\boldmath
$\chi$}}^{\{\sigma \rho \}}+  \label{yukk} \\
&&+\overline{L}_{f}^{\alpha }R_{df}^{\beta }\overline{H}_{d}[A_{df}\cdot 
\overline{\mbox{\boldmath $\eta$}}_{[\alpha \sigma ]}\overline{%
\mbox{\boldmath $\eta$}}_{[\beta \rho ]}+B_{df}\cdot \epsilon _{\alpha \beta
\sigma }\mbox{\boldmath $\eta$}_{\rho }{\bf S}^{l}]\overline{%
\mbox{\boldmath
$\chi$}}^{\{\sigma \rho \}}+  \nonumber \\
&&+N_{R}^{\alpha }N_{R}^{\beta }[A_{NN}\cdot \overline{%
\mbox{\boldmath
$\eta$}}_{[\alpha \sigma ]}\overline{\mbox{\boldmath $\eta$}}_{[\beta \rho
]}]\overline{\mbox{\boldmath $\chi$}}^{\{\sigma \rho \}}  \nonumber
\end{eqnarray}
for the up and down fermions, quarks ($f=q$) and leptons ($f=l$), and
right-handed neutrinos. They contain only one triplet scalar $%
\mbox{\boldmath $\eta$}_{\alpha }$ (and its SUSY counterpart $\overline{%
\mbox{\boldmath $\eta$}}_{[\alpha \beta ]}\equiv \epsilon _{\alpha \beta
\gamma }\overline{\mbox{\boldmath $\eta$}}^{\gamma }$) developing VEVs (\ref
{comp}) along both the first and third directions at the same time. They
also include one of the sextets $\overline{\mbox{\boldmath $\chi$}}%
^{\{\alpha \beta \}}$, so as to be properly arranged in the family space. We
remark that the couplings $W_{Y1}$ (\ref{yukkk}), like the general Yukawa
couplings (\ref{yuk'}), use horizontal triplets for their symmetric parts
and anti-triplets for the antisymmetric parts. However the couplings $W_{Y2}$
(\ref{yukk}), on the contrary, prefer to use anti-triplets for the symmetric
parts and triplets for the antisymmetric parts. Just those couplings
collected in $W_{Y2}$ turn out to correspond, as we will see later, to the
minimal mixing of quarks and leptons with the proportionality condition (\ref
{prop}) between their mass matrix elements. Note that the couplings in $%
W_{Y2}$ include their own hierarchy making singlet scalar field ${\bf S}$.
This scalar ${\bf S}$ determines (see Eq.(\ref{small})) the small components
in the VEVs of the triplets $\eta $ ($\overline{\eta }$) and $\zeta $ ($%
\overline{\zeta }$) and thereby underlie the basic hierarchy between the
diagonal elements $M_{22}$ and $M_{33}$ in the quark and lepton mass
matrices (see later). Thus, it looks natural that the hierarchy in their
off-diagonal elements is also dependent solely on the power ($l$) of the
same scalar field ${\bf S}$ and not on the other scalar ${\bf I}$ introduced
in the couplings $W_{Y1}$ (\ref{yukkk}). Here again all the horizontal
scalar fields \mbox{\boldmath  $\eta$}, $\overline{\mbox{\boldmath $\eta$}}$%
, $\overline{\mbox{\boldmath $\chi$}}$ and ${\bf S}$ are scaled by the $%
SU(3)_{F}$ symmetry mass scale parameter $M_{F}$, so that the coupling
constants $A_{uf,df,NN}$ and $B_{uf,df}$ are all dimensionless and assumed
to be of order unity. According to the $U(1)$ symmetry, this hierarchy
should be the same for all the fermions involved in $W_{Y2}$. The symmetric
couplings containing the bi-linears of the $\mbox{\boldmath $\eta$} $ scalar
give masses to the fermions of the heaviest third family and have no
suppression, apart from the contributions coming from the small components
in its VEV configuration (\ref{comp}). For simplicity, we also propose that
the up-down hierarchy in the Yukawa couplings (\ref{yukk}) is entirely given
by the ratio of the VEVs of the ordinary Higgs doublets $H_{u} $ and $%
\overline{H}_{d}$ of the MSSM, $\tan \beta =v_{u}/v_{d}$, thus considering
the large $\tan \beta $ case. One can, of course, include instead extra
powers of the singlet scalar fields ${\bf S}$ in the Yukawa couplings of the
down fermions to generate the up-down mass hierarchy.

The Yukawa couplings (\ref{yukk}), when taken for all the quarks and leptons
involved, correspond in fact to the combined scenario A+A$^{*}$ for flavour
mixing. In this case there is only one (natural) hierarchy parameter $y_{S}=%
\frac{<S>}{M_{F}}$ for all the fermion mass-matrices. However, in the
combined scenario B+B$^{*}$ where the masses and mixings of the up quarks
are determined by the Yukawa couplings $W_{Y1}$ (\ref{yukkk}), while those
of the down quarks and leptons are determined by the Yukawa couplings $%
W_{Y2} $ (\ref{yukk}), generally two independent hierarchy parameters $y_{I}=%
\frac{<I>}{M_{F}}$ and $y_{S}=\frac{<S>}{M_{F}}$ appear in the model. In
general, one can hardly expect that the different Yukawa superpotentials, $%
W_{Y1}$ and $W_{Y2}$, should lead to a similar hierarchy pattern for the
corresponding fermions. We shall assume (see section 3.3.4) that the
hierarchy making scalars $I$ and $S$ develop VEVs which are close in value,
leading to close values of the hierarchy parameters $y_{I}$ and $y_{S}$.
However they must have quite different $U(1)$ charges, in order to provide
the required power-like hierarchy in the quark and lepton mass matrices.

The couplings $W_{Y1}$ (\ref{yukkk}) and $W_{Y2}$ (\ref{yukk}) are the only
Yukawa couplings which give masses to the quarks and leptons. Other possible 
$SU(3)_{F}$ couplings containing the same fermion and scalar superfields
are, in fact, disallowed by the extra $U(1)$ symmetry mentioned above (see
section 3.3.4 for more details).

\subsubsection{ Mass-matrices of quarks and leptons}

We show now that, from the four presently (experimentally) allowed LFMG
scenarios for the combined flavour mixing of quarks and leptons, the $%
SU(3)_{F}$ theory admits only one; namely the B+B$^{*}$ scenario which,
fortunately, seems to be the most preferable with regard to the experimental
situation in quark mixing (the present value of the $V_{ub}$ element, see
section 2.1).

Towards this end let us first consider the Yukawa couplings $W_{Y2}$ (\ref
{yukk}) to understand why they, while being good for all the other fermions,
cannot work for the up quarks and right-handed neutrinos. Substituting all
the VEV values of the horizontal triplets (\ref{comp}) and sextets (\ref
{svevs}), as well as those of the ordinary MSSM doublets $H_{u}$ and $%
\overline{H}_{d}$, into the Yukawa couplings $W_{Y2}$ (\ref{yukk}), one
obtains the following mass matrices for the quarks and leptons:

\begin{eqnarray}
M_{uf} &=&m_{u}^{0}\left( 
\begin{array}{lll}
A_{uf}\overline{{\bf x}}^{2} & 0 & A_{uf}\overline{{\bf x}} \overline{{\bf X}%
} \\ 
0 & A_{uf}\overline{{\bf x}}^{2} & B_{uf}{\bf XS}^{l} \\ 
A_{uf}\overline{{\bf x}}\overline{{\bf X}} & -B_{uf}{\bf XS}^{l} & A_{uf} 
\overline{{\bf X}}^{2}
\end{array}
\right) ,  \label{up} \\
&&  \nonumber \\
M_{df} &=&m_{d}^{0}\left( 
\begin{array}{lll}
A_{df}\overline{{\bf x}}^{2} & 0 & A_{df}\overline{{\bf x}} \overline{{\bf X}%
} \\ 
0 & A_{df}\overline{{\bf x}}^{2} & B_{df}{\bf XS}^{l} \\ 
A_{df}\overline{{\bf x}}\overline{{\bf X}} & -B_{df}{\bf XS}^{l} & A_{df} 
\overline{{\bf X}}^{2}
\end{array}
\right)  \label{down}
\end{eqnarray}

\begin{equation}
M_{NN}=m_{NN}^{0}\left( 
\begin{array}{lll}
A_{NN}\overline{{\bf x}}^{2} & 0 & A_{NN}\overline{{\bf x}}\overline{{\bf X}}
\\ 
0 & A_{NN}\overline{{\bf x}}^{2} & 0 \\ 
A_{NN}\overline{{\bf x}}\overline{{\bf X}} & 0 & A_{NN}\overline{{\bf X}}^{2}
\end{array}
\right)  \label{maj}
\end{equation}
Here the mass parameters $m_{u}^{0},$ $m_{d}^{0}$ and $m_{NN}^{0}$ include
all the other VEV factors appearing with the Yukawa couplings in $W_{Y2}$ :

\begin{equation}
m_{u}^{0}=<H_{u}>\overline{\mbox{\boldmath $\chi$}},\quad m_{d}^{0}=<%
\overline{H}_{d}>\overline{\mbox{\boldmath $\chi$}},\quad m_{NN}^{0}=M_{F}%
\overline{\mbox{\boldmath $\chi$}}  \label{m}
\end{equation}
(the bold letters stand everywhere for the properly scaled VEVs, for example 
${\bf X\equiv }X/M_{F}$ etc). Now, after diagonalization of the 1-3 blocks
in the mass matrices $M_{uf}$, $M_{df}$ and $M_{NN}$, one immediately comes
to the LFMG form (\ref{LFM1}) for the mass-matrices of all the fermions
involved (specifying those for the up and down quarks, neutrinos, charged
leptons and right-handed neutrinos, respectively):

\begin{eqnarray}
M_{U} &=&m_{u}^{0}\left( 
\begin{array}{lll}
0 & B_{U}\frac{\overline{{\bf x}}}{\overline{{\bf X}}}{\bf XS}^{l} & 0 \\ 
-B_{U}\frac{\overline{{\bf x}}}{\overline{{\bf X}}}{\bf XS}^{l} & A_{U} 
\overline{{\bf x}}^{2} & B_{U}{\bf XS}^{l} \\ 
0 & -B_{U}{\bf XS}^{l} & A_{U}\overline{{\bf X}}^{2}
\end{array}
\right) ,  \label{up'} \\
&&  \nonumber \\
M_{D} &=&m_{d}^{0}\left( 
\begin{array}{lll}
0 & B_{D}\frac{\overline{{\bf x}}}{\overline{{\bf X}}}{\bf XS}^{l} & 0 \\ 
-B_{D}\frac{\overline{{\bf x}}}{\overline{{\bf X}}}{\bf XS}^{l} & A_{D} 
\overline{{\bf x}}^{2} & B_{D}{\bf XS}^{l} \\ 
0 & -B_{D}{\bf XS}^{l} & A_{D}\overline{{\bf X}}^{2}
\end{array}
\right) ,  \label{down'} \\
&&  \nonumber \\
M_{N} &=&m_{u}^{0}\left( 
\begin{array}{lll}
0 & B_{N}\frac{\overline{{\bf x}}}{\overline{{\bf X}}}{\bf XS}^{l} & 0 \\ 
-B_{N}\frac{\overline{{\bf x}}}{\overline{{\bf X}}}{\bf XS}^{l} & A_{N} 
\overline{{\bf x}}^{2} & B_{N}{\bf XS}^{l} \\ 
0 & -B_{N}{\bf XS}^{l} & A_{N}\overline{{\bf X}}^{2}
\end{array}
\right) ,  \label{neu} \\
&&  \nonumber \\
M_{E} &=&m_{d}^{0}\left( 
\begin{array}{lll}
0 & B_{E}\frac{\overline{{\bf x}}}{\overline{{\bf X}}}{\bf XS}^{l} & 0 \\ 
-B_{E}\frac{\overline{{\bf x}}}{\overline{{\bf X}}}{\bf XS}^{l} & A_{E} 
\overline{{\bf x}}^{2} & B_{E}{\bf XS}^{l} \\ 
0 & -B_{E}{\bf XS}^{l} & A_{E}\overline{{\bf X}}^{2}
\end{array}
\right)  \label{lep} \\
&&  \nonumber \\
M_{NN} &=&m_{NN}^{0}\left( 
\begin{array}{lll}
0 & 0 & 0 \\ 
0 & A_{NN}\overline{{\bf x}}^{2} & 0 \\ 
0 & 0 & A_{NN}\overline{{\bf X}}^{2}
\end{array}
\right)  \label{NN}
\end{eqnarray}
with the desired proportionality condition (\ref{prop}) or (\ref{ABab})
between their diagonal and off-diagonal elements. This diagonalization, as
one can readily see, is reached by a physically irrelevant rotation (being
the same for the up and down fermions) with a hierarchically small angle $%
\Theta_{13}^{u,d}\simeq \frac{\overline{{\bf x}}} {\overline{{\bf X}}} =%
\frac{\overline{x}}{\overline{X}}\approx (\frac{S}{M_{F}})^{r}$ (see (\ref
{small}))\footnote{%
We also neglected the small addition of the order of $O(S^{2})$ to the new
(33)-elements appearing in all these mass-matrices after the diagonalization
of the 1-3 blocks.}.

This is, in fact, a crucial point for deriving the LFMG picture within the $%
SU(3)_{F}$ symmetry framework. On the one hand, the scalar $\eta $ in the
Yukawa couplings $W_{Y2}$ (\ref{yukk}) must develop VEVs along the first and
third directions in family space, in order that the proportionality
condition (\ref{ABab}) be satisfied. On the other, these VEVs are allowed to
induce at most only non-physical extra elements in the mass-matrices $M_{uf}$
and $M_{df}$ (\ref{up}, \ref{down}) that could be rotated away by some
rotation with $\Theta _{13}^{u}=\Theta _{13}^{d}$. Otherwise, one would have
large physical mixing of the order of $O(y)$ (where $y\approx 0.1$ is the
hierarchy parameter introduced in our discussion of neutrino oscillations in
section 2.2.1) between the first and third families of fermions that is
actually excluded, at least for quarks where this mixing angle is in fact of
order of $O(y^{3})$, as follows from the $V_{ub}$ element value in the
observed CKM matrix (see (\ref{ckm})).

There seems to be no other way of getting the LFMG in the $SU(3)_{F}$ model,
when the same form $W_{Y2}$ (\ref{yukk}) is taken for the Yukawa couplings
of both the up and down fermions . If so, however, all the matrices $M_{uf}$%
, $M_{df}$ and $M_{NN}$ appear, as one can readily see from their explicit
forms (\ref{up'}-\ref{NN}), to be largely proportional to each other. This
results in an equality of the mass ratios for the heavier leptons and quarks

\begin{equation}
\frac{m_{c}}{m_{t}}\simeq \frac{m_{s}}{m_{b}}\simeq \frac{m_{\mu }} {m_{\tau
}}\simeq \frac{M_{N2}}{M_{N3}}\simeq \frac{M_{NN2}}{M_{NN3}}  \label{ratio}
\end{equation}
which should hold with an accuracy of a few percent (given in fact by the
ratios of the masses of the first and second families). While such an
equality approximately works for the down quarks and charged leptons and
also might be taken for the Dirac masses of neutrinos, it is certainly
unacceptable for the up quarks \cite{data}. This means that we are left with
just one possible scenario B for the minimal quark mixing in the $SU(3)_{F}$
context, according to which the up quarks follow the pattern of direct (1-3)
mixing (the alternative II, see general discussion in section 2.2), while
the down quark mixing is given by a mass matrix of the above type (\ref
{down'}) corresponding to the alternative I.

As to the lepton mixing, both of the experimentally allowed scenarios A$^{*}$
and B$^{*}$ (section 2.2) follow the alternative I for the Dirac
mass-matrices of leptons. Thus $M_N$ and $M_E$ should have the form given
above (\ref{neu}, \ref{lep}), resulting in the proportionality condition (%
\ref{ratio}) between the Dirac masses of neutrinos and charged leptons. This
condition should be necessarily kept into account when analyzing neutrino
oscillations (as we have done in section 2.2.), since it allows us to
predict the hierarchy parameter $y$ appearing in the neutrino Dirac mass
matrix: $y=\sqrt{m_{\mu }/m_{\tau }}$. If we take (as we actually did) that
the neutrino Majorana mass matrix also depends on the same parameter $y$
then all the characteristic feature of the neutrino oscillations can be
predicted according to Eq.~(\ref{pred1}) in scenario A$^{*}$ and
Eq.~(\ref{pred2}) in scenario B$^{*}$.

However, scenario A$^{*}$ presupposes the stronger hierarchy (\ref{h2})
between the second and third eigenvalues of the neutrino Majorana mass
matrix $M_{NN}$ to get the observed large mixings (see section 2.2) than the
hierarchy (\ref{h1}) appearing in the Dirac mass matrix $M_{N}$ of
neutrinos. The requirement for such a hierarchy to be in the matrix $M_{NN}$
is certainly in conflict with the mass proportionality condition (\ref{ratio}%
), which appears in scenario A$^{*}$ within the $SU(3)_{F}$ framework.
Another drawback of this scenario A$^{*}$ in the $SU(3)_{F}$ framework is
the absence of the antisymmetric mixing term in the Yukawa couplings for the
right-handed neutrinos (see (\ref{yukk})) that leads to one massless
Majorana state in $M_{NN}$ (\ref{NN}). Thus, scenario A$^{*}$ of lepton
flavour mixing should be dropped in favour of scenario B$^{*}$, with direct
(1-3) mixing in the matrix $M_{NN}$ just as it appears for the up quarks in
scenario B (see section 3.3.5 for further discussion of the matrix $M_{NN}$%
). So, we can conclude that the combined scenario B+B$^{*}$ is in substance
the only one admitted in the $SU(3)_{F}$ theory. For this case the mass
proportionality condition (\ref{ratio}) takes the reduced form:

\begin{equation}
\frac{m_{s}}{m_{b}}\simeq \frac{m_{\mu }}{m_{\tau }}\simeq \frac{M_{N2}}{%
M_{N3}}  \label{ratio'}
\end{equation}
which seem to work better than the whole equation (\ref{ratio}), however,
not yet in a quite satisfactory way. Actually, this relation between the
masses of the down quarks and leptons is similar to the usual GUT relation,
while arising in a different way. However, by enlarging the set of
horizontal scalars, one can rearrange or completely decouple the masses of
the quarks and leptons.

Now in scenario B, after family symmetry breaking according to the VEVs of
the horizontal triplets (\ref{comp}), one is immediately led from the Yukawa
couplings collected in $W_{Y1}$ to the mass matrix for the up quarks:

\begin{equation}
M_{U}=m_{U}^{0}\left( 
\begin{array}{lll}
A_{U1}{\bf z}^{2} & 0 & A_{U1}{\bf zZ}-B_{U}\overline{{\bf Y}}{\bf I}^{n} \\ 
0 & A_{U2}{\bf Y}^{2}{\bf I^{k}} & 0 \\ 
A_{U1}{\bf zZ}+B_{U}\overline{{\bf Y}}{\bf I}^{n} & 0 & A_{U1}{\bf Z}^{2}
\end{array}
\right)  \label{up''}
\end{equation}
where the mass factor $m_{U}^{0}$ is given by the electroweak symmetry scale
involved

\begin{equation}
m_{U}^{0}=<H_{u}>  \label{sc}
\end{equation}
Further, after diagonalization of the symmetric terms in the 1-3 block, one
comes to the familiar form

\begin{equation}
M_{U}=m_{U}^{0}\left( 
\begin{array}{lll}
0 & 0 & -B_{U}\overline{{\bf Y}}{\bf I}^{n} \\ 
0 & A_{U2}{\bf Y}^{2}{\bf I}^{k} & 0 \\ 
B_{U}\overline{{\bf Y}}{\bf I}^{n} & 0 & A_{U1}{\bf Z}^{2}
\end{array}
\right)  \label{up'''}
\end{equation}
discussed in section 2 (see (\ref{LFM2})).

Let us note that the above partial diagonalization of the mass matrix (\ref
{up''}) was in fact reached by a rotation for the up quarks with the
hierarchically small angle $\Theta_{13}^{U}\simeq -\frac{{\bf z}}{{\bf Z}}=%
\frac{z}{Z}\approx (\frac{S}{M_{F}})^{r}$ (see (\ref{small})). It is crucial
for the scenario B+B$^{*}$ considered that this rotation turns out to be
physically irrelevant: it is in fact the same as for the down quarks and
leptons, despite the latter having mass-matrices of another type (\ref{down}%
) or (\ref{down'}). However, the orthogonality (\ref{ort}) of the VEVs of
the scalars involved keeps the rotations in both sectors strictly equivalent
to one another

%\medskip
\begin{equation}
\overline{\eta }^{\alpha }\zeta _{\alpha }=0\rightarrow \frac{\overline{{\bf %
x}}}{\overline{{\bf X}}}+\frac{{\bf z}}{{\bf Z}}=0\rightarrow \Theta
_{13}^{D,N,E}=\Theta _{13}^{U}  \label{ort'}
\end{equation}
As a result, one can simultaneously transform the primary (``unrotated'')
matrices for the up quarks (\ref{up''}), on the one hand, and those of the
down quarks (\ref{down}) and leptons (neutrinos (\ref{up}) and charged
leptons \ref{down})), on the other, into the final (``rotated'') matrices (%
\ref{up'''}) and (\ref{down'}-\ref{lep}) respectively. These matrices are
just those which exactly correspond to the combined scenario B+B$^{*}$ for
flavour mixing of quarks and leptons. These matrices follow, as we have seen
above, from the Yukawa couplings $W_{Y1}$ (\ref{yukkk}) and $W_{Y2}$ (\ref
{yukk}) when the family symmetry $SU(3)_{F}$ spontaneously breaks.

\subsubsection{Hierarchies in masses and mixings}

We saw in the previous sections that the extra $U(1)$ symmetry, requiring
the same powers $k$, $n$ and $l$ of the hierarchy making scalars $I$ and $S$
in similar Yukawa couplings of quarks and leptons, leads in the considered
scenario B+B$^{*}$ to similar hierarchical mass matrices for the down quarks
(\ref{down'}) and the leptons (\ref{neu}, \ref{lep}). As a result, we
obtained the mass relations (\ref{ratio'}) for the quarks and leptons in the
scenario B+B$^{*}$.

However, besides this general conclusion the $U(1)$ symmetry turns out, as
we now show, to generate the required hierarchy in the mass matrices of the
quarks and leptons. In order to show this in the most general form, we start
with a scale-invariant superpotential which, in its only difference with the
above-used Higgs superpotential $W$ (\ref{spot}), contains no mass parameter
for any of the scalar superfields, triplets and sextets, involved. Instead,
we assume that their masses are forbidden by the $U(1)$ symmetry and are
generated by the VEV of some singlet scalar field $T$ from the following
allowed couplings:

%\begin{mathletters}
\begin{equation}
W_{T}=T(A_{\eta }\eta \overline{\eta }+A_{\xi }\xi \overline{\xi }+A_{\zeta
}\zeta \overline{\zeta }+A_{\chi }\chi \overline{\chi }+A_{\omega }\omega 
\overline{\omega })+P(I,S,T)  \label{T}
\end{equation}
So the masses of the scalars, which were directly introduced above (see
section 3.3.1), are now given by

%\end{mathletters}
\begin{equation}
M_{\eta ,\xi ,\zeta ,\chi ,\omega }=A_{\eta ,\xi ,\zeta ,\chi ,\omega }\cdot
<T>  \label{M}
\end{equation}
through the VEV of the scalar field $T$ and the coupling constants $A_{\eta
,\xi ,\zeta ,\chi ,\omega }$. Thus, the VEV of the scalar field $T$ gives
the basic flavour scale $M_{F}$ in the model, which is proposed now to
appear dynamically rather than being introduced by hand. This, in substance,
should stem from the polynomial $P(I,S,T)$ which includes all three singlet
scalars; one of which ($T$) gives the basic mass scale in the model, while
the other two ($I$ and $S$) determine the mass and mixing hierarchy of the
quarks and leptons. The hierarchy parameters are basically given by the
characteristic VEV ratios $I/T$ and $S/T$ determined by the proper minimum
of the polynomial $P(I,S,T)$ (see below).

Based on the above comments, we can now write the new total Higgs
superpotential (see Eqs.~(\ref{spot}, \ref{spot'}, \ref{acc}, \ref{T})) in
the form

\begin{eqnarray}
W_{tot} &=&T(A_{\eta }\eta \overline{\eta }+A_{\xi }\xi \overline{\xi }%
+A_{\zeta }\zeta \overline{\zeta }+A_{\chi }\chi \overline{\chi }+A_{\omega
}\omega \overline{\omega })+  \nonumber \\
&&+f\cdot \epsilon ^{\alpha \beta \gamma }\eta _{\alpha }\xi _{\beta }\zeta
_{_{\gamma }}+\overline{f}\cdot \epsilon _{\alpha \beta \gamma }\overline{%
\eta }^{\alpha }\overline{\xi }^{\beta }\overline{\zeta }^{_{\gamma }}+ 
\nonumber \\
&&+F\cdot \epsilon ^{\alpha \beta \gamma }\epsilon ^{\alpha ^{\prime }\beta
^{\prime }\gamma ^{\prime }}\chi _{\{\alpha \alpha ^{\prime }\}}\chi
_{\{\beta \beta ^{\prime }\}}\omega _{\{\gamma \gamma ^{\prime }\}}+%
\overline{F}\cdot \epsilon _{\alpha \beta \gamma }\epsilon _{\alpha ^{\prime
}\beta ^{\prime }\gamma ^{\prime }}\overline{\chi }^{\{\alpha \alpha
^{\prime }\}}\overline{\chi }^{\{\beta \beta ^{\prime }\}}\overline{\omega }%
^{\{\gamma \gamma ^{\prime }\}}+  \nonumber \\
&&+a\cdot \overline{\eta }\zeta {\bf S}^{r}M_{F}+b\cdot \omega \overline{%
\zeta }\overline{\zeta }{\bf S}^{p}  \label{snew}
\end{eqnarray}
with ${\bf I}$ and ${\bf S}$ standing for $I/M_{F}$ and $S/M_{F}$ and the
polynomial $P(I,S,T)$ being as yet omitted. One can now quickly derive, from
the new total Higgs superpotential $W_{tot}$ and the Yukawa couplings $%
W_{Y1} $ (\ref{yukkk}) and $W_{Y2}$ (\ref{yukk}), that the $U(1)$ charges of
all the horizontal scalar fields involved are given as follows in terms of
the charges of the singlet scalars $S$ and $I$:

\begin{eqnarray}
Q_{\eta } &=&-\frac{5l}{3}Q_{S}+\frac{8n-2k}{3}Q_{I},  \nonumber \\
\quad Q_{\xi } &=&-\frac{2l}{3}Q_{S}+\frac{5n-2k}{3}Q_{I},\qquad  \nonumber
\\
Q_{\zeta } &=&-\frac{2l}{3}Q_{S}+\frac{10n-k}{6}Q_{I},  \nonumber \\
Q_{\chi } &=&-\frac{4l+3p}{3}Q_{S}+\frac{14n-5k}{6}Q_{I},  \nonumber \\
Q_{\omega } &=&-\frac{l+3p}{3}Q_{S}+\frac{8n+k}{6}Q_{I},  \nonumber \\
Q_{T} &=&-lQ_{S}+\frac{4n-k}{2}Q_{I}  \label{charges}
\end{eqnarray}
The $U(1)$ charges for their supersymmetric counterparts $\overline{\eta }$, 
$\overline{\xi }$, $\overline{\zeta }$, $\overline{\chi }$ and $\overline{%
\omega }$ are given by

\begin{equation}
Q_{\overline{\eta },\overline{\xi },\overline{\zeta },\overline{\chi },%
\overline{\omega }}=-Q_{\eta ,\xi ,\zeta ,\chi ,\omega }+2Q_{T}  \label{ch'}
\end{equation}
In fact, the $U(1)$ charges of all the horizontal scalars can be expressed
in terms of one independent charge \ $Q_{S}$, as one can readily see from
the relation

\begin{equation}
Q_{I}=\frac{2l+r}{3n-k}Q_{S}  \label{QQ}
\end{equation}
arising from the $\overline{\eta }\zeta $ mixing term in the total
superpotential $W_{tot}$. Here the numbers $k$, $l$ and $n$ are just the
powers of the singlet scalars $I$ and $S$ in the Yukawa couplings $W_{Y1}$
and $W_{Y2}$, while the numbers $r$ and $p$ appear as the powers of $S$ in
the triplet-triplet and triplet-sextet mixing terms respectively in the
superpotential $W_{tot}$ (\ref{snew}).

When the above horizontal superfields undergo a $U(1)$ transformation,
according to the charges (\ref{charges}), the total superpotential $W_{tot}$
acquires a physically inessential overall phase\footnote{%
This means that the $U(1)$ considered is in fact the global $R$-symmetry $%
U(1)_{R}$ of the superpotential $W_{tot}$ (\ref{snew}).}

\begin{equation}
W_{tot}\rightarrow e^{i3\alpha _{T}}W_{tot}  \label{W}
\end{equation}
where $\alpha _{T}$ is the transformation phase corresponding to the singlet
scalar superfield $T$. Under the same transformation, the Yukawa couplings $%
W_{Y1}$ (\ref{yukkk}) and $W_{Y2}$ (\ref{yukk}) are left invariant.

As to the matter superfields, the $U(1)$ charges of the quarks

\begin{equation}
\left( 
\begin{array}{c}
U_{L} \\ 
D_{L}
\end{array}
\right) _{\alpha }, \quad U_{R}^{\alpha }, \quad D_{R}^{\alpha }  \label{q}
\end{equation}
leptons

\begin{equation}
\left( 
\begin{array}{c}
N_{L} \\ 
E_{L}
\end{array}
\right) _{\alpha },\quad N_{R}^{\alpha },\quad E_{R}^{\alpha }  \label{l}
\end{equation}
(see (\ref{left}, \ref{right})) and the ordinary Higgs doublets, $H_{u}$ and 
$\overline{H}_{d}$, are in substance arbitrary\footnote{%
One can immediately check that, in the B+ B$^{*}$ scenario, the Yukawa
couplings $W_{Y1}$ (\ref{yukkk}) and $W_{Y2}$ (\ref{yukk}) only give five
equations for the eight charges (or phases) of the six independent fermion (%
\ref{q}, \ref{l}) and two Higgs ($H_{u}$, $\overline{H}_{d}$) multiplets
involved.} and can be chosen in many possible ways.

We will now investigate which powers $k$, $l$, $n$ and $r$ are required to
get the observed mass and mixing hierarchy from the quark and lepton mass
matrices. First of all one should make sure that no other matrix elements
are allowed, by the $U(1)$ symmetry and the chosen powers, than those
required by the scenario B+B$^{*}$ in all four mass matrices considered;
namely for the up quarks ($M_{U}$ (\ref{up'''})), down quarks ($M_{D}$ \ref
{down'})), neutrinos ($M_{N}$ (\ref{neu})) and charged leptons ($M_{E}$ (\ref
{lep})). One can readily check that for an even value of $r$ and an odd
value of $k$ with

\begin{equation}
n>k \qquad (k \ \mbox{is odd})  \label{nk}
\end{equation}
the undesired horizontal scalar combinations of type $\eta \eta $, $\eta \xi 
$, $\eta \zeta $ and $\xi \zeta $ (accompanied by any powers of the singlet
scalars $I$ and $S$) in the Yukawa couplings $W_{Y1}$ (\ref{yukkk}), as well
as the combinations $\overline{\zeta }\overline{\zeta }\overline{\chi }$, $%
\overline{\xi }\overline{\xi }\overline{\chi }$ and $\zeta \overline{\chi }$
in the Yukawa couplings $W_{Y2}$ (\ref{yukk}) are strictly prohibited to
appear. This is important since their presence might crucially destroy
scenario B+B$^{*}$. The only allowed extra combination is $\xi \overline{%
\chi }$ which appears in $W_{Y2}$, thus giving an additional Yukawa coupling

\begin{equation}
\overline{L}_{f}^{\alpha }R_{df}^{\beta }\overline{H}_{d}[B_{df}^{\prime
}\cdot \epsilon _{\alpha \beta \sigma }\mbox{\boldmath $\xi$}_{\rho } {\bf I}%
^{n}]\overline{\mbox{\boldmath $\chi$ }}^{\{\sigma \rho \}}  \label{ex}
\end{equation}
This actually turns out to be negligibly small as compared to the other
couplings for the down quarks and leptons ($f=D,N,E$) involved. So, one can
see that the form of all the mass matrices considered are highly protected
by the $U(1)$ symmetry from any scalar contributions, other than those
allowed by the B+B$^{*}$ scenario of the minimal mixing of quarks and
leptons.

Let us consider now the mass matrices $M_{D}$, $M_{N}$ and $M_{E}$ (\ref
{down'}-\ref{lep}). All of them have a similar structure with the
proportionality (\ref{prop}) between their diagonal and off-diagonal matrix
elements. The natural orders of magnitude of these mass matrix elements are
given by powers of the VEV of the singlet scalar $S$ scaled by the basic
mass $M_{F}$ ($y=\frac{<S>}{M_{F}}$):

\begin{equation}
(M_{f})_{33}\approx m^{0}, \ (M_{f})_{22}\approx m^{0}\overline{{\bf x}}%
^{2}\approx m^{0}y^{2r}, \ (M_{f})_{23}\approx m^{0}y^{l},\
(M_{f})_{12}\approx m^{0}y^{l+r}  \label{H1}
\end{equation}
($f=D,N,E$) as directly follows from equations (\ref{down'}-\ref{lep}) and (%
\ref{small}). Since for the down quarks the approximate equality $%
(M_{D})_{22}\approx (M_{D})_{23}$ (or equivalently $m_s \approx \sqrt{m_d m_b%
}$) is known to work well phenomenologically, one is led to the following
relation between the powers $l$ and $r$:

\begin{equation}
l=2r  \label{p}
\end{equation}
Thus, according to the scenario B+B$^{*}$ considered, all three matrices $%
M_{D}$, $M_{N}$ and $M_{E}$ should have a similar hierarchical structure of
the above type (\ref{H1}) with $l=2r$ where $r$ is as yet an arbitrary even
number.

We turn now to the mass matrix $M_{U}$ (\ref{up'''}) for the up quarks. The
natural orders of magnitude of its elements are given by the appropriate
powers of the VEV of another singlet $I$ (scaled by $M_{F}$). Taking for
simplicity $y_{I}\approx y_{S}$ $\equiv y$ ($<I>$ $\approx $ $<S>$), one
readily finds for the matrix elements of $M_{U}$:

\begin{equation}
(M_{U})_{33}\approx m_{U}^{0},\quad (M_{U})_{22}\approx m_{U}^{0}y^{k},\quad
(M_{U})_{13}\approx m_{U}^{0}y^{n}  \label{H1'}
\end{equation}
For consistency with the phenomenological observation \cite{data} concerning
the quark masses of the second and third families, $(m_{c}/m_{t})^{2}\approx
(m_{s}/m_{b})^{3}$, comparison of the diagonal matrix elements of $M_{U}$
and $M_{D}$ gives, for the case of an even value for $r$ and an odd value of 
$k$ (see eq.~(\ref{nk})), a relation of the type 
\begin{equation}
k=3r+1  \label{k}
\end{equation}
And, finally, according to the inequality (\ref{nk}), the minimal choice for
the power $n$ is

\begin{equation}
n=3r+2  \label{nr}
\end{equation}
One can see that the above relations for the powers $k$, $l$ and $n$ lead to
approximately the same hierarchy in the mass matrices of quarks and leptons
for any (even) value of the number $r$. However, we will proceed further
with $r=2$, thus finally establishing the values

\begin{equation}
k=7, \quad l=4, \quad n=8 \quad (r=2)  \label{values}
\end{equation}
for the other powers.

According to these values, the quark and lepton mass matrices (\ref{down'}-%
\ref{lep}) and (\ref{up'''}) in the scenario B+B$^{*}$ acquire the
hierarchical forms:

\begin{equation}
M_{f}={\bf m}_{f}\left( 
\begin{array}{lll}
0 & \lambda \rho _{f}y^{6} & 0 \\ 
-\lambda \rho _{f}y^{6} & \lambda ^{2}y^{4} & \rho _{f}y^{4} \\ 
0 & -\rho _{f}y^{4} & 1
\end{array}
\right) ,\quad f=D,N,E  \label{H2}
\end{equation}
and

\begin{equation}
M_{U}={\bf m}_{U}\left( 
\begin{array}{lll}
0 & 0 & \sigma y^{8} \\ 
0 & \alpha y^{7} & 0 \\ 
\sigma y^{8} & 0 & 1
\end{array}
\right)   \label{H2'}
\end{equation}
Here the parameters $\rho _{f}$ , $\alpha $ and $\sigma $, being ratios of
Yukawa coupling constants, are all proposed to be of order unity, while the
mass scale factors ${\bf m}_{f}$ and ${\bf m}_{U}$ are as yet ``unrotated''
masses of the heaviest third family of fermions involved. Also, we have
parameterized the common VEV ratio $\frac{\overline{x}}{\overline{X}}$
contained in the matrices $M_{D,N,E}$ as $\frac{\overline{x}}{\overline{X}}%
=\lambda y^{2}$ ($\lambda \approx 1$), according to the behaviour of the
small components of the VEVs appearing on the horizontal triplets $\eta $
and $\zeta $ (see (\ref{small})). Note that, in the matrices $M_{f}$ (\ref
{H2}), we have not included the negligibly small matrix element $%
(M_{f})_{13}=O(y^{8})$ stemming generally from the ``extra'' Yukawa coupling
(\ref{ex})\footnote{%
This element is, as one can readily see from matrix (\ref{H2}), of order of
the lightest mass $m_{d}$ for down quarks. Being inessential in any other
respect it will, however, contribute to the $V_{ub}$ element, thus 
changing the formula (\ref{angles}) given in section 2 
to a formula of the type $\left| V_{ub}\right|
\simeq s_{13}\simeq \left| \sqrt{\frac{m_{u}}{m_{t}}}+e^{i\gamma }\frac{m_{d}%
}{m_{b}}\right| $. In the maximal CP violation case (see subsection 3.3.6), 
$\gamma =\frac{\pi }{2}$, this new contribution turns out
to be about 10 percent of the main one, and is neglected below.}.

One can see that, due to the high values of the powers in equation (\ref
{values}), even a very smooth starting hierarchy\footnote{%
To avoid confusion we should note that the hierarchy parameter $y$
introduced here differs from that used in section 2.2 (and denoted by the
same letter): the old $y$ appears to be approximately equal to the square of
the new one.} $y\approx (m_{s}/m_{b})^{1/4}\approx 0.4$ leads, as is
apparent in the mass matrices (\ref{H2}) and (\ref{H2'}), to the observed
hierarchy in the masses and mixings of the quarks and leptons. Apart from
the values of the weak mixing angles (\ref{angles}) which, as was already
shown in section 2.1, are in a good agreement with experiment, there appear
characteristic relations between the quark (and lepton) masses of the type:

\begin{equation}
\frac{m_{u}}{m_{c}}\approx \left(\frac{m_{c}}{m_{t}}\right)^{9/7},\ \frac{%
m_{d}}{m_{s}} \approx \frac{m_{s}}{m_{b}},\ \frac{m_{u}}{m_{c}} \approx
\left(\frac{m_{d}}{m_{s}}\right)^{9/4},\ \frac{m_{c}}{m_{t}} \approx \left(%
\frac{m_{s}}{m_{b}}\right)^{7/4}  \label{rel}
\end{equation}
which seem to be successfull phenomenologically. Actually, one needs to know
only the heaviest mass for each family of fermions---the other masses are
then given in terms of this mass and the hierarchy parameter $y$. At the
same time it is well to bear in mind that all these mass relations appear at
the flavour scale $M_{F}$, which could be as high as the GUT scale or even
the Planck scale. Doing so, one then needs to project these relations down
to laboratory energies to compare them with experiment \cite{albr}.

We consider finally the form of the polynomial $P(I,S,T)$ introduced above
(see Eq.~(\ref{T}) and how the VEVs of the basic singlet scalars $I$, $S$
and $T$ are determined. Note first of all that it is not immediately
evident, for the $U(1)$ charges of the scalars $I$, $S$ and $T$ already
determined from $W_{tot}$, $W_{Y1}$ and $W_{Y2}$ (see Eqs. (\ref{charges}, 
\ref{QQ})) and from the hierarchy requirements (\ref{p},\ref{k},\ref{nr}),
that such a non-trivial $U(1)$ invariant polynomial exists. Remarkably,
however, just for these charges of the three singlet scalars $I$, $S$ and $T$
a polynomial of the type

\begin{equation}
P(I,S,T)=a_{ABC}T^{A}I^{B}S^{C}  \label{P}
\end{equation}
actually exists, where the summation over all the allowed integer powers $A$%
, $B$ and $C$ is imposed\footnote{%
Non-integer or negative values are of course not allowed for the powers in
the Higgs superpotential or in the Yukawa couplings, due to their generic
analyticity.}. This polynomial determines the basic mass scale $M_{F}\simeq $
$<T>$ in the model, as well as the characteristic hierarchies related with
the two other VEVs $<I>$ and $<S>$ (all the high-order terms in $P$ are
scaled by appropriate powers of the mass $M_{F}$ contained in the coupling
constants $a_{ABC}$).

Actually, in order for a term in this invariant polynomial to exist, the $%
U(1)$ symmetry condition

\begin{equation}
AQ_{T}+BQ_{I}+CQ_{S}=3Q_{T}
\end{equation}
for the charges of the scalars $I$, $S$ and $T$ should be satisfied.
Substitution of the values of these charges, expressed (see (Eqs.~\ref
{charges}, \ref{QQ}, \ref{p}, \ref{k}, \ref{nr})) in terms of the power $r$
of the scalar $S$ in the total Higgs superpotential $W_{tot}$ (\ref{snew}),
in this condition gives the equation

\begin{equation}
\frac{A-3}{2}(7r+5)+\frac{5}{3}B+C\frac{6r+5}{3r}=0
\end{equation}
There are only a finite number of positive integer solutions of this
equation for the powers $A$, $B$ and $C$. For the case $r=2$ taken in our
model, only the following four terms appear in $P(I,S,T)$

\begin{equation}
P(I,S,T)=a_{300}T^{3}+a_{241}T^{2}I^{4}S+a_{182}T^{1}I^{8}S^{2}+
a_{0123}I^{12}S^{3}  \label{PPP}
\end{equation}
while not disturbing the other parts of the total Higgs superpotential $%
W_{tot}$ (\ref{snew}). In this case one can show from the corresponding
F-term equations (together with those stemming from $W_{tot}$) that in
general non-zero VEVs appear for all the singlet scalars $I$, $S$ and $T$.
Even a very smooth hierarchy between them could then lead, as was argued
above, to the observed hierarchy in masses and mixings of the quarks and
leptons.

\subsubsection{Neutrino masses and oscillations}

We have not yet considered the masses of the right-handed neutrinos and the
Majorana mass matrix $M_{NN}$ needed to construct the effective mass matrix $%
M_{\nu }$ for the ordinary neutrinos (\ref{nu}), using the see-saw mechanism 
\cite{GRS}. According to the B+B$^{*}$ scenario, the right-handed neutrinos
should have a mass matrix similar to that of the up quarks, having symmetric
Yukawa couplings with the same structure as $W_{Y1}$ (\ref{yukkk}), but the
antisymmetric couplings are necessarily absent for Majorana neutrinos. As a
result, with only the symmetric Yukawa couplings contained in $W_{Y1}$, one
is unavoidably led to one massless right-handed neutrino. Therefore, some
additional symmetric coupling is actually needed\footnote{%
One might imagine introducing a symmetric mixing term containing the
non-diagonal horizontal triplet combination $\eta \zeta $ in the Yukawa
coupling $W_{Y1}$ (\ref{yukkk}). However, as we have seen in the above, such
a term is strictly prohibited by the $U(1)$ symmetry.}. Fortunately such a
coupling, generating right-handed neutrino masses, automatically appears in $%
W_{Y1}$ once the triplet-sextet coupling term $\Delta W_{2}$ (\ref{acc}) is
included in the total superpotential $W_{tot}$ (\ref{snew}). One can then
immediately confirm that the horizontal sextet $\omega $, with the proper $%
U(1)$ charge value given in (\ref{charges}), is allowed to contribute to $%
W_{Y1}$ together with the triplet pairs $\zeta \zeta $ and $\xi \xi $. Thus,
the total Yukawa superpotential for the right-handed neutrinos reads as

\begin{equation}
W_{Y11}=N_{R}^{\alpha }N_{R}^{\beta }[A_{NN0}\cdot \omega _{\{\alpha \beta
\}}{\bf S}^{p}{\bf T+}A_{NN1}\cdot \mbox{\boldmath$ \zeta$}_{\alpha }%
\mbox{\boldmath $\zeta$}_{\beta }+A_{NN2}\cdot \mbox{\boldmath $\xi$}%
_{\alpha }\mbox{\boldmath $\xi$}_{\beta }{\bf I}^{k}]  \label{y'}
\end{equation}
The first term in $W_{Y11}$ should, of course, also be included in the
Yukawa couplings $W_{Y1}$ (\ref{yukkk}) for the up quarks. However, this
term, while being crucial for neutrino mass formation, turns out to be
inessential for the up quarks (for the high value of the power $p$ of the
singlet $S$ chosen below).

After substituting the VEVs of the sextet $\omega _{\alpha \beta }$ (\ref
{svevs}) and the triplets $\zeta $ and $\xi $ (\ref{comp}, \ref{small}) into
the Yukawa couplings $W_{Y11}$ (\ref{y'}), one immediately comes to the
Majorana mass matrix for the right-handed neutrinos:

\begin{equation}
M_{NN}=M_{F}\left( 
\begin{array}{lll}
A_{NN1}{\bf z}^{2} & 0 & A_{NN1}{\bf zZ} \\ 
0 & A_{NN2}{\bf Y}^{2}{\bf I^{k}} & 0 \\ 
A_{NN1}{\bf zZ} & 0 & A_{NN1}{\bf Z}^{2}+A_{NN0}{\bf \omega S}^{p}{\bf T}
\end{array}
\right)  \label{Mnn}
\end{equation}
This matrix can be written in a convenient hierarchical form, like the other
mass matrices (\ref{H2}, \ref{H2'}), using the already established hierarchy
indices (see ~(\ref{values})) and keeping the power $p$ as yet undefined:

\begin{equation}
M_{NN}={\bf m}_{NN}\left( 
\begin{array}{lll}
y^{4} & 0 & y^{2} \\ 
0 & \gamma y^{7} & 0 \\ 
y^{2} & 0 & 1+\beta y^{p}
\end{array}
\right)  \label{H2''}
\end{equation}
Here ${\bf m}_{NN}$, $\beta $ and $\gamma $ are appropriate combinations of
the masses and coupling constants in (\ref{Mnn}) and the value ${\bf T\equiv 
}\frac{T}{M_{F}}\simeq 1$ has been used. The physical masses of the Majorana
neutrinos are then given by the equations

\begin{equation}
M_{NN1}\simeq {\bf m}_{NN}\beta y^{4+p},\quad M_{NN2}\simeq {\bf m}%
_{NN}\gamma y^{7},\quad M_{NN3}\simeq {\bf m}_{NN}  \label{Nmasses}
\end{equation}

The mass matrix $M_{\nu }$ for the physical neutrinos can now readily be
constructed according to the see-saw formula (\ref{nu}), with the neutrino
Dirac mass matrix $M_{N}$ (\ref{H2}) taken in its ``unrotated'' form (\ref
{up})

\begin{equation}
M_{N}={\bf m}_{N}\left( 
\begin{array}{lll}
\lambda ^{2}y^{4} & 0 & \lambda y^{2} \\ 
0 & \lambda ^{2}y^{4} & \rho _{N}y^{4} \\ 
\lambda y^{2} & -\rho _{N}y^{4} & 1
\end{array}
\right)  \label{unrot}
\end{equation}
and the Majorana mass matrix $M_{NN}$ given above (\ref{H2''}). The inverse
of the Majorana mass matrix $M_{NN}$ is readily formed 
\begin{equation}
M_{NN}^{-1}=\frac{1}{\beta y^{4+p}{\bf m}_{NN}}\left( 
\begin{array}{lll}
1+\beta y^{p} & 0 & -y^{2} \\ 
0 & \beta \gamma ^{-1}y^{-3+p} & 0 \\ 
-y^{2} & 0 & y^{4}
\end{array}
\right) ,  \label{inv}
\end{equation}
and gives rise to an effective light neutrino mass matrix $M_{\nu }$. To
simplify its form we take, as in the general phenomenological case (see 
section 2.2), for all the order-one parameters $\lambda $, $\rho _{N}$ 
and $\beta $ contained in the matrices $M_{N}$ and $M_{NN}$, an extra 
condition of the type (\ref{natur}): 
\begin{equation}
\left| \Delta -1\right| \compoundrel<\over\sim y^{4}\qquad (\Delta \equiv
\lambda, \rho_{N}, \beta )  \label{naturprime}
\end{equation}
according to which they are supposed to be equal to
unity with a few percent accuracy$^{11}$. After this simplification and the
choice $p=8$, the matrix $M_{\nu }$ takes a transparent symmetrical form
(ignoring some inessential higher order terms)

\begin{equation}
M_{\nu }={\bf m}_{\nu }\left( 
\begin{array}{lll}
y^{4} & -y^{2} & -y^{2} \\ 
-y^{2} & 1+\gamma y^{-1} & 1+\gamma y^{-1}\delta  \\ 
-y^{2} & 1+\gamma \delta y^{-1} & 1+\gamma y^{-1}(1+\delta ^{2})
\end{array}
\right)   \label{Mnu}
\end{equation}
where the matrix $M_{\nu }$ has been first constructed from matrices  $M_{N}$
(\ref{unrot}) and $M_{NN}^{-1}$ (\ref{inv}) according to the see-saw formula
(\ref{nu}) and then brought to the appropriate physical basis, by making a
rotation in the (1-3) block (equal to that already applied to the matrix of
the charged leptons $M_{E}$ so as to bring it to the form (\ref{H2})). Here
the mass parameter

\begin{equation}
{\bf m}_{\nu }=-\frac{{\bf m}_{N}^{2}}{{\bf m}_{NN}}\gamma ^{-1}y
\label{nu3}
\end{equation}
essentially corresponds to the mass of the heaviest neutrino, while the
parameter $\delta \equiv \frac{\lambda -1}{y^{4}}$ satisfies $\left| \delta
\right| \leq 1$ according to the condition (\ref{naturprime}) above.

One can see that the matrix $M_{\nu }$ is crucially dependent on the
parameter $\gamma $ related to the triplet-triplet term $%
\mbox{\boldmath
$\xi$}\mbox{\boldmath $\xi$}{\bf I}^{k}$ in the Yukawa superpotential $%
W_{Y11}$ (\ref{y'}). While all the dimensionless coupling constants in $%
W_{Y11}$ are proposed (just as for the other Yukawa and Higgs coupling
constants in the model) to be of the natural order $O(1)$, the parameter $%
\gamma $ should be considerably smaller to have an experimentally acceptable
mass interval between the masses of the second and third family neutrinos.
In particular, if we take $\gamma \approx y^{4}$ (which is just of the order 
$O(m_{s}/m_{b})$, see the matrices (\ref{H2})), we come to an attractive
picture for neutrino masses and oscillations. Actually, after
diagonalization of the matrix $M_{\nu }$, one can readily see that the
ratios of the neutrino masses are\footnote{%
For simplicity, we take $\delta =1$ here. For the other limiting case, $%
\delta =0$, the mass ratios are a little changed, $m_{\nu 1}:m_{\nu
2}:m_{\nu 3}=\frac{1}{4}y^{7}:\frac{1}{2}y^{3}:1$.}

\begin{equation}
m_{\nu 1}:m_{\nu 2}:m_{\nu 3}=\frac{1}{2}y^{7}:\frac{1}{4}y^{3}:1
\label{rat}
\end{equation}
and the small-mixing angle MSW solution (SMA) for neutrino oscillations
naturally emerges, with the following predictions for the basic parameters:

\begin{equation}
\sin ^{2}2\theta _{atm}\simeq 1,\quad \sin ^{2}2\theta _{sun}%
\compoundrel<\over\sim 2y^{4},\quad U_{e3}\simeq \frac{y^{2}}{\sqrt{2}}%
,\quad \frac{\Delta m_{sun}^{2}}{\Delta m_{atm}^{2}}\simeq \frac{y^{6}}{16}
\label{pred3}
\end{equation}
Here, in deriving the limit on the contribution to $\sin ^{2}2\theta _{sun}$
from the neutral lepton sector, we have used the 
inequality (\ref{naturprime}) for the $(M_{\nu })_{12}$ element 
obtained after successive (2-3) and (1-3) block diagonalizations.

For the phenomenological value of the hierarchy parameter in our model, $y
\approx (m_{s}/m_{b})^{1/4} \approx 0.4$, the predictions (\ref{pred3}) turn
out to be inside of the experimentally allowed intervals for the SMA
solution (\ref{sma}). Note that for the case when the above parameters $%
\lambda $, $\rho _{N}$ and $\beta $ are strictly equal to unity the (1-2)
mixing completely disappears, while the (2-3) mixing determining the value
of $\sin^{2}2\theta _{atm}$ is left maximal. Note also that in the SMA case
one must expect, as we mentioned above (see section 2.2), a sizable
contribution to $\sin ^{2}2\theta _{sun}$ stemming from the charged lepton
sector (see Eqs.(\ref{charged})). Remarkably, this contribution taken alone

\begin{equation}
\sin ^{2}2\theta _{sun}\simeq 4\sin ^{2}\theta _{e\mu }=4\frac{m_{e}}{m_{\mu
}}  \label{ch}
\end{equation}
gives a good approximation to the SMA solution.

So, one can conclude that the $SU(3)_{F}$ theory of flavour successfully
describes the neutrino masses and mixings as well. Remarkably, just the
neutrino oscillation phenomena already observed could give the first real
hint as to where the scale $M_{F}$ of the $SU(3)_{F}$ flavour symmetry might
be located. Actually, if the right-handed neutrinos are not sterile under $%
SU(3)_{F}$ (as they are under all other gauge symmetries) then their masses
should be directly related to the flavour scale, just like the masses of the
quarks and leptons are related to the electroweak scale. We treat the
right-handed neutrinos, like the other quarks and leptons, as triplets $%
N_{R}^{\alpha }$ under $SU(3)_{F}$ (see (\ref{right})) which acquire their
masses from the $SU(3)_{F}$ invariant couplings $W_{Y11}$ (\ref{y'}), when
the horizontal scalars develop VEVs and flavour symmetry breaks. Due to the
chiral nature of $SU(3)_{F}$, there appear no mass terms for the Majorana
neutrinos $N_{R}^{\alpha }$ when the symmetry is unbroken. Thus, it is
apparent that the mass of the heaviest Majorana neutrino, appearing from the
Yukawa coupling $W_{Y11}$, should be of the same order as the flavour scale $%
M_{F}$, since the basic Yukawa coupling constants are all proposed to be of
order unity. Using now equation (\ref{nu3}), with the heaviest Dirac
neutrino mass ${\bf m}_{N}$ taken as large as the top quark mass $m_{t}$,
the heaviest physical neutrino mass ${\bf m}_{\nu }$ determined from the
experimentally observed value of $\Delta m_{atm}^{2}$ \cite{superkam}, ${\bf %
m}_{\nu }\simeq \sqrt{\Delta m_{atm}^{2}}\simeq 0.07$ eV and the hierarchy
parameter $y\approx (m_{s}/m_{b})^{1/4}$, one comes to an approximate value
for the flavour scale

\begin{equation}
M_{F}\approx {\bf m}_{NN}=-\frac{{\bf m}_{N}^{2}}{{\bf m}_{\nu }}\gamma
^{-1}y\simeq \frac{m_{t}^{2}}{{\bf m}_{\nu }}\left( \frac{m_{b}}{m_{s}}%
\right) ^{3/4}\simeq 10^{16}\ \mbox{GeV}  \label{scale}
\end{equation}
from which the masses and mixings of the quarks and leptons are basically
generated. Due to the enhancement factor $\gamma ^{-1}y$ appearing in the $%
SU(3)_{F}$ model considered, this scale turns out to be as large as the
well-known SUSY GUT scale \cite{GD}. The high scale $M_{F}$ of the flavour
symmetry found above (see (\ref{scale})) would mean that the $SU(3)_{F}$
flavour symmetry could be part of some extended SUSY GUT, either in a direct
product form like $SU(5)\otimes SU(3)_{F}$ \cite{jon}, $SO(10)\otimes
SU(3)_{F}$ \cite{wu} and $E(6)\otimes SU(3)_{F}$ \cite{wu} or even in the
form of a family unifying simple group like $SU(8)$ GUT \cite{jon2}.

\subsubsection{Spontaneous CP violation}

It seems that the whole picture of flavour mixing of quarks and leptons,
including the hierarchical pattern of the masses and mixing angles, can
arise as a consequence of the spontaneous breakdown of the family symmetry $%
SU(3)_{F}$ considered. We show, in this section, that even CP violation can
be spontaneously induced in the $SU(3)_{F}$ theory of flavour. Furthermore,
when it occurs, one is unavoidably led to maximal CP violation.

The essential point is that, in general, the horizontal scalars (\ref{set})
develop complex VEVs (\ref{comp}, \ref{svevs}) leading to CP violation, even
if all the coupling constants in the superpotential $W_{tot}$ (\ref{snew})
and the Yukawa couplings in $W_{Y1}$ (\ref{yukkk}) and $W_{Y2}$ (\ref{yukk})
are taken to be real. Let us consider this in more detail. For simplicity,
one can always choose real VEVs for the singlet scalars $I$, $S$ and $T$,
among the possible VEV solutions stemming from the invariant polynomial $%
P(I,S,T)$ (\ref{P}). Then, as follows from the triplet VEV equations (\ref
{tvevs}, \ref{small}), the phases of the large and small components (\ref
{comp}) of their VEVs should satisfy the conditions

\begin{eqnarray}
\delta _{X}+\delta _{\overline{X}} &=&0, \quad \delta _{Y}+ \delta _{%
\overline{Y}}=0, \quad \delta _{Z}+\delta _{\overline{Z}}=0, \quad \delta
_{X}+\delta _{Y}+\delta _{Z}=\pi \quad  \label{ph1} \\
\delta _{\overline{x}} &=&\pi -\delta _{X}+\delta _{Y}+2\delta _{Z}, \quad
\delta _{z}=\pi -\delta _{X}+2\delta _{Z}, \quad \delta _{\overline{z}}=\pi
-\delta _{X}  \nonumber
\end{eqnarray}
There is one more non-trivial condition, stemming from the imaginary part of
the triplet F-term equations (\ref{ftrms}) containing also the singlet
scalar $S$. When one uses the previous conditions (\ref{ph1}), this extra
condition reads as follows:

\begin{equation}
-\delta _{X}+\delta _{Y}+3\delta _{Z}=\pi \cdot {\bf k},\quad {\bf k}=0,1
\label{ph2}
\end{equation}
Note now that, due to the exact $U(1)$ symmetry of the total superpotential $%
W_{tot}$ (\ref{snew}) and the polynomial $P(I,S,T)$ (\ref{P}), one can
always choose the phase of the $U(1)$ transformation in such a way as to
make one of the phases $\delta _{X}$, $\delta _{Y}$ or $\delta _{Z}$ in the
equations (\ref{ph1}, \ref{ph2}) vanish. Thus, taking $\delta _{Y}=0$, one
finally comes to two non-trivial solutions for the phases of the large and
small components of the triplet VEVs:

\begin{eqnarray}
\delta _{X} &=&\pi /4, \quad \delta _{Z}=3\pi /4,  \nonumber \\
\delta _{\overline{x}} &=&\pi /4, \quad \delta _{z}=\pi /4, \quad \delta _{%
\overline{z}}=3\pi /4  \label{ph3}
\end{eqnarray}
and

\begin{eqnarray}
\delta _{X} &=&\pi /2, \quad \delta _{Z}=\pi /2,  \nonumber \\
\delta _{\overline{x}} &=&3\pi /2, \quad \delta _{z}=3\pi /2, \quad \delta _{%
\overline{z}}=\pi /2  \label{ph4}
\end{eqnarray}
depending on the two possible values of ${\bf k}$, ${\bf k}=0$ and ${\bf k}%
=1 $ respectively. We show below by applying these triplet VEV phase values
to the quark and lepton mass matrices that, while the second solution (\ref
{ph4}) is CP-conserving, the first one (\ref{ph3}) leads to maximal
CP-violation. There are no other solutions for the phases of the triplet
VEVs in the $SU(3)_{F}$ model considered\footnote{%
As to the VEVs of the horizontal sextets $\chi $ and $\omega $ they always
can be chosen real. Actually, by a proper choice of the phase of the $%
SU(3)_{F}$ transformations (particularly its $\lambda ^{8}$ subgroup
transformation) one can take the phase of $<\chi >$ to be zero. Then, as
follows from the sextet F-term equations (\ref{sex}), the phase of $<\omega
> $ turns out to vanish as well.}.

Let us consider the CP-conserving case (\ref{ph4}) first. As one can quickly
confirm, the mass matrices $M_{U}$ (\ref{up'''}) and $M_{D}$ (\ref{down'})
of the up and down quarks, in this case containing the phases of the
horizontal triplet VEVs given by equation (\ref{ph4}), are strictly
Hermitian with real diagonal elements and pure imaginary off-diagonal ones.
Such matrices $M_{U}$ and $M_{D}$, with five and three texture zero
structure respectively, cannot lead to CP violation---the phase of the
corresponding Jarlskog determinant \cite{jarl} vanishes identically.

By contrast, the CP-violating case (\ref{ph3}) results in non-Hermitian mass
matrices $M_{U}$ and $M_{D}$. One can find their explicit form, in terms of
the masses of the up and down quarks and the triplet VEV phases (\ref{ph3}),
by forming their Hermitian products $M_{U}M_{U}^{+}$ and $M_{D}M_{D}^{+}$
and then using the conservation of their traces, determinants and sums of
principal minors. Doing so, we find for the matrices $M_{U}$ and $M_{D}$:

\begin{equation}
M_{U}=\left( 
\begin{array}{lll}
0 & 0 & -\sqrt{m_{u}m_{t}}e^{i\pi /2} \\ 
0 & m_{c}e^{i\pi /2} & 0 \\ 
\sqrt{m_{u}m_{t}}e^{i\pi /2} & 0 & m_{t}-m_{u}
\end{array}
\right)  \label{Ufin}
\end{equation}
and

\begin{equation}
M_{D}=\left( 
\begin{array}{lll}
0 & -\sqrt{m_{d}m_{s}}e^{-i3\pi /4} & 0 \\ 
\sqrt{m_{d}m_{s}}e^{-i3\pi /4} & m_{s} & \sqrt{m_{d}m_{b}}e^{i3\pi /4} \\ 
0 & -\sqrt{m_{d}m_{b}}e^{i3\pi /4} & m_{b}-m_{d}
\end{array}
\right)  \label{Dfin}
\end{equation}
where we have removed the inessential common phase factors.

Notably, while the diagonalization of the non-Hermitian matrices $M_{U}$ and 
$M_{D}$ requires different unitary transformations for the left-handed and
right-handed quarks, the Hermitian constructions $M_{U}M_{U}^{+}$ and $%
M_{D}M_{D}^{+}$ are diagonalized by unitary transformations acting on just
the left-handed quarks. These are precisely the transformations which
contribute to the weak interaction mixings of quarks collected in the CKM
matrix. Thus, one must diagonalize the $M_{U}M_{U}^{+}$ and $M_{D}M_{D}^{+}$
first and then construct the CKM matrix. Proceeding in such a manner, we
immediately find that the diagonalization of $M_{U}M_{U}^{+}$ and $%
M_{D}M_{D}^{+}$ is carried out by unitary matrices of the type

\begin{equation}
V_{U}=R_{13}^{U}\Phi ^{U},\qquad V_{D}=R_{12}^{D}R_{23}^{D}\Phi ^{D}
\label{V2}
\end{equation}
They in fact contain the same plane rotations $R_{13}^{U}$ ($s_{13}\simeq 
\sqrt{m_{u}/m_{t}}$), $R_{12}^{D}$ ($s_{12}\simeq \sqrt{m_{d}/m_{s}}$) and $%
R_{23}^{D}$ ($s_{23}\simeq \sqrt{m_{d}/m_{b}}$) as those used for the
diagonalization of the phenomenological Hermitian mass matrices $M_{U}$ (\ref
{LFM2}) and $M_{D}$ (\ref{LFM1}) considered in section 2.2. However, the
phase matrices $\Phi ^{U}$ and $\Phi ^{D}$ now become

\begin{equation}
\Phi ^{U}=\left( 
\begin{array}{lll}
e^{i\pi /2} & 0 & 0 \\ 
0 & 1 & 0 \\ 
0 & 0 & 1
\end{array}
\right) , \quad \Phi ^{D}=\left( 
\begin{array}{lll}
1 & 0 & 0 \\ 
0 & e^{-i3\pi /4} & 0 \\ 
0 & 0 & 1
\end{array}
\right)  \label{phi}
\end{equation}
Finally, substituting all the above rotations and phases into the CKM matrix 
$V_{CKM}=V_{U}V_{D}^{\dagger }$ (\ref{CKM1}) and properly rephasing the $c$
quark field ($c\rightarrow e^{i3\pi /4}c$), one comes to

\begin{equation}
V_{CKM}=\left( 
\begin{array}{lll}
c_{12}c_{13}-is_{12}s_{13}s_{23} & -s_{12}c_{13}-is_{13}s_{23}c_{12} & 
-is_{13}c_{23} \\ 
s_{12}c_{23} & c_{12}c_{23} & -s_{23} \\ 
-s_{13}c_{12}-is_{12}s_{23}c_{13} & s_{13}s_{12}-is_{23}c_{12}c_{13} & 
-ic_{13}c_{23}
\end{array}
\right)  \label{CKM3}
\end{equation}
While this is in substance a new parameterization for the CKM matrix, it is
quite close to the standard parameterization \cite{data}, as one can show by
comparing the moduli of the matrix elements in them both. In such a way, one
readily concludes that the CP violation phase in $V_{CKM}$ (\ref{CKM3}) is
in fact maximal, $\delta =\pi /2$, while all the mixing angles are rather
small being basically determined by the masses of the lightest quarks $u$
and $d$ (see Eqs.~(\ref{angles})), as was expected.

\section{Conclusions}

The present observational status of quark flavour mixing, as described by
the CKM matrix elements \cite{data}, shows that the third family $t$ and $b$
quarks are largely decoupled from the lighter families. At first sight, it
looks quite surprising that not only the 1-3 ``far neighbour'' mixing
(giving the $V_{ub}$ element in $V_{CKM}$) but also the 2-3 ``nearest
neighbour'' mixing ($V_{cb}$) happen to be small compared with the
``ordinary'' 1-2 Cabibbo mixing ($V_{us}$) which is determined, according to
common belief, by the lightest $u$ and $d$ quarks. This led us to the idea
that all the other mixings, and primarily the 2-3 mixing, could also be
controlled by the masses $m_{u}$ and $m_{d}$ and that the above-mentioned
decoupling of the third family $t$ and $b$ quarks is determined by the
square roots of the corresponding mass ratios $\sqrt{\frac{m_{u}}{m_{t}}}$
and $\sqrt{\frac{m_{d}}{m_{b}}}$ respectively. So, in the chiral symmetry
limit $m_{u}=m_{d}=0$, not only does CP violation vanish, as argued in \cite
{review}, but all the flavour mixings disappear as well.

In such a way the Lightest Family Mass Generation (LFMG) mechanism for
flavour mixing of quarks was formulated in section 2.1, with two possible
scenarios A and B. We found that the LFMG mechanism reproduces well the
values of the already well measured CKM matrix elements and gives
distinctive predictions for the yet poorly known ones, in both scenarios A
and B. One could say that, for the first time, there are compact working
formulae (especially compact in scenario B) for all the CKM angles in terms
of quark mass ratios. The only unknown parameter is the CP violating phase $%
\delta $. Taking it to be maximal ($\delta =\frac{\pi }{2}$), we obtain a
full determination of the CKM matrix that is numerically summarized in
section 2.1. The LFMG mechanism was extended to the lepton sector in section
2.2, again with two possible scenarios A$^{*}$ and B$^{*}$. These were shown
to be quite successful phenomenologically as well, naturally leading to a
consistent description of the presently observed situation in solar and
atmospheric neutrino oscillations . Remarkably, the same mechanism results
simultaneously in small quark and large lepton mixing.

{}From the theoretical point of view, the basic LFMG mechanism arises from
the generic proportionality condition (\ref{prop}) between diagonal and
off-diagonal elements of the mass matrices. One might think that this
condition suggests some underlying flavour symmetry, probably a non-abelian $%
SU(N)$ symmetry, treating the N families in a special way. Indeed, for $N=3$
families, we have found that the local $SU(3)_{F}$ chiral family (or
horizontal) symmetry \cite{jon}, considered in detail in section 3, seems to
be a good candidate. The $SU(3)_{F}$ family symmetry is accompanied by an
additional global $U(1)$ symmetry, which seems to be very important for the
recovery of the final picture of flavour mixing of quarks and leptons. There
is a whole plethora of $SU(3)_{F}$ invariant Yukawa couplings available.
However, due to the extra protecting symmetry $U(1)$, one is able to choose
them in a special form which leads to the minimal flavour mixing of quarks
and leptons. There are four pairs of experimentally allowed scenarios (see
section 2) for the combined quark and lepton flavour mixings: A+A$^{*}$, A+ B%
$^{*}$, B+A$^{*}$ and/or B+B$^{*}$. We found that the $SU(3)_{F}$ theory
only admits the scenario B+B$^{*}$ which, fortunately, seems to provide the
best fit to the experimental data on quark mixing. It also leads to the SMA
solution for the solar neutrino problem. Besides its crucial role in
selecting the pattern of flavour mixing among the four scenarios involved,
the $U(1)$ symmetry turns out, as we have shown, to determine the form of
the hierarchy in the quark and lepton mass matrices\footnote{%
Note, however, that we had to take the parameter $\gamma $ in the Majorana
mass matrix~(\ref{H2''}) to be small ($\gamma \approx y^{4}$), while all the
parameters in the quark and lepton mass matrices are supposed to of the
order $O(1)$.}. Actually, one only needs to know the heaviest mass for each
family of fermions---the other masses are then given in terms of this mass 
and the input hierarchy parameter $y$.

The high scale $M_{F}$ of the flavour symmetry found here from neutrino
oscillations means that the local flavour symmetry $SU(3)_{F}$ might be
interpreted as part of some extended SUSY GUT; this could appear in a direct
product form like $SU(5)\otimes SU(3)_{F}$, $SO(10)\otimes SU(3)_{F}$ and $%
E(6)\otimes SU(3)_{F}$ or even in the form of a simple group like the $SU(8)$
GUT unifying all families.

And the last point, which is especially noteworthy, is that the
symmetry-breaking horizontal scalar fields, triplets and sextets of $SU(3)$,
in general develop complex VEVs and, in cases linked to the LFMG mechanism,
transmit a maximal CP violating phase $\delta =\frac{\pi }{2}$ to the
effective Yukawa couplings involved. Apart from the direct predictability of 
$\delta $, the possibility that CP symmetry is broken spontaneously like
other fundamental symmetries of the Standard Model seems very
attractive---both aesthetically and because it gives some clue to the origin
of maximal CP violation in the CKM matrix.

So, an $SU(3)$ family symmetry seems to be a good candidate for the basic
theory underlying our proposed LFMG mechanism, although we do not exclude
the possibility of other interpretations as well. Certainly, even without a
theoretical derivation of Eq.~(\ref{prop}), the LFMG mechanism can be
considered as a successful predictive ansatz in its own right. Its further
testing could shed light on the underlying flavour dynamics and the way
towards the final theory of flavour.

\section*{Acknowledgements}

We should like to thank H. Fritzsch, H. Leutwyler, D. Sutherland and Z. -Z.
Xing for interesting discussions. JLC is grateful to PPARC for financial
support (PPARC grant No. PPA/V/S/2000/) during his visit to Glasgow
University in June-July of 2000, when part of this work was done. Financial
support to JLC from INTAS grant No. 96-155 and the Joint Project grant from
the Royal Society are also gratefully acknowledged. CDF and HBN would like
to thank the EU commission for financial support from grants No. SCI-0430-C
(PSTS) and No. CHRX-CT-94-0621. Also JLC, CDF and HBN would like to
acknowledge financial support from INTAS grant No. 95-567.

\end{document}